\documentclass[twocolumn]{aastex63}
\usepackage[caption=false]{subfig}
\usepackage{color}

\usepackage{xcolor}
\usepackage[normalem]{ulem}

\newcommand{\lya}{Ly$\alpha$}

\accepted{October 15, 2021}

\shorttitle{Morphology and Kinematics of a \lya-selected Nebula}
\shortauthors{Sanderson et al.}

\begin{document}

\title{Mapping the Morphology and Kinematics of a \lya-selected Nebula at z=3.15 with MUSE}

\author[0000-0001-5825-7683]{Kelly N. Sanderson}
\affiliation{Department of Astronomy, New Mexico State University, P.O.Box 30001, MSC 4500, Las Cruces, NM, 88033, USA}

\author[0000-0001-8302-0565]{Moire M. K. Prescott}
\affiliation{Department of Astronomy, New Mexico State University, P.O.Box 30001, MSC 4500, Las Cruces, NM, 88033, USA}

\author[0000-0001-8415-7547]{Lise Christensen}
\affiliation{Cosmic Dawn Center (DAWN), Copenhagen, Denmark}
\affiliation{Niels Bohr Institute, University of Copenhagen, Jagtvej 128, 2200 Copenhagen N, Denmark}

\author[0000-0002-8149-8298]{Johan Fynbo}
\affiliation{Cosmic Dawn Center (DAWN), Copenhagen, Denmark}
\affiliation{Niels Bohr Institute, University of Copenhagen, Jagtvej 128, 2200 Copenhagen N, Denmark}

\author[0000-0002-9994-505X]{Palle M{\o}ller}
\affiliation{European Southern Observatory, Karl-Schwarzschildstrasse 2, D-85748 Garching bei M\"unchen, Germany}
\affiliation{Niels Bohr Institute, University of Copenhagen, Jagtvej 128, 2200 Copenhagen N, Denmark}

\begin{abstract}

Recent wide-field integral field spectroscopy has revealed the detailed properties of high redshift \lya\ nebulae, most often targeted due to the presence of an active galactic nucleus (AGN). Here, we use VLT/MUSE to resolve the morphology and kinematics of a nebula initially identified due to strong \lya\ emission at $z\sim3.2$ \citep[LABn06;][]{Nilsson2006}. Our observations reveal a two-lobed \lya\ nebula, at least $\sim$173 pkpc in diameter, with a light-weighted centroid near a mid-infrared source (within $\approx$17.2 pkpc) that appears to host an obscured AGN. The \lya\ emission near the AGN is also coincident in velocity with the kinematic center of the nebula, suggesting that the nebula is both morphologically and kinematically centered on the AGN. Compared to AGN-selected \lya\ nebulae, the surface brightness profile of this nebula follows a typical exponential profile at large radii ($>$25 pkpc), although at small radii, the profile shows an unusual dip at the location of the AGN. The kinematics and asymmetry are similar to, and the CIV and HeII upper limits are consistent with other AGN-powered \lya\ nebulae. Double-peaked and asymmetric line profiles suggest that \lya\ resonant scattering may be important in this nebula. These results support the picture of the AGN being responsible for powering a \lya\ nebula that is oriented roughly in the plane of the sky. Further observations will explore whether the central surface brightness depression is indicative of either an unusual gas or dust distribution or variation in the ionizing output of the AGN over time.

\end{abstract}


\section{Introduction} 

Recent studies have revealed extended diffuse \lya\ emission that covers nearly 100\% of the sky around high redshift galaxies \citep{Wisotzki2018, Leclercq2020}. The mechanisms responsible for lighting up the cosmic web via \lya\ radiation have been debated for a long time. Processes such as fluorescence powered by Active Galactic Nuclei \citep[AGN; e.g.,][]{Cantalupo2005, Kollmeier2010, Kimock2020}, resonance scattering of centrally produced \lya\ photons \citep[e.g.,][]{Verhamme2006, Steidel2011}, shock-heating in powerful galactic winds \citep[e.g.,][]{Taniguchi2000, Taniguchi2001,Mori2004}, and gravitational cooling \citep[e.g.,][]{Rosdahl2012} of collisionally excited HI atoms have all been put forth to explain the observed emission. In some cases multiple powering mechanisms are believed to contribute to the observed emission \citep[e.g.][]{Vanzella2017, Herenz2020}; in others, one mechanism appears to dominate \citep[e.g.][]{Borisova2016, Battaia2018}.

While many studies of diffuse \lya\ emission have focused on \lya-emitting galaxies \citep[LAEs;][]{Steidel2000, Rauch2008} or \lya-emitting halos centered on AGN \citep[e.g.][]{Heckman1991, Christensen2006, Villar-Martin2007, Cantalupo2012, Cantalupo2014, Borisova2016}, long-standing questions remain about the nature and powering mechanisms of the most dramatic \lya-emitting regions in the Universe, i.e. giant \lya\ nebulae;  \citep[or \lya\ ``blobs,'' LAB; e.g.,][]{Francis1996,Steidel2000,Matsuda2004,Dey2005, Prescott2015a, Cai2017, Cantalupo2014}. These nebulae show a range of unusual morphologies and in some cases trace out filaments connecting individual galaxies \citep[][]{Erb2011, Battaia2019}. In general, they do appear to reside preferentially in overdense regions \citep[e.g.,][]{Steidel2000, Prescott2009,Yang2009}, but  the frequent lack of obvious ionizing sources in or near many of these nebulae makes their energetics difficult to interpret \citep[but see also][]{Overzier2013}.

Being able to resolve the environments in which \lya\ nebulae reside and the kinematics of the emitting gas will provide a better opportunity to pinpoint which powering mechanisms contribute to the overall emission. Narrowband (NB) and broadband imaging \citep[e.g.,][]{Fynbo1999, Steidel2000, Matsuda2004, Yang2009, Prescott2012a, Prescott2012b} combined with follow-up long slit spectroscopy \citep[e.g.,][]{Weidinger2005, Matsuda2006, Yang2009, Zafar2011, Prescott2015a} have been useful in constraining the nebula morphologies and in investigating the kinematics along selected dimensions through the emitting gas, but a key difficulty is not being able to fully map the gas dynamics of the entire nebula with respect to the associated galaxy population. Integral field spectrograph (IFS) observations provide a spatially resolved view of this gas \citep[e.g.,][]{Borisova2016, North2017, Vanzella2017, Battaia2018, Herenz2020} and allow us to search for signatures of either in-flowing or out-flowing material, potentially providing evidence for cold accretion flows or feedback processes, respectively. IFS data can also be used to detect other emission lines in addition to \lya\ that can further constrain the dominant powering source.

Using IFS data from the VLT’s MUSE \citep{Bacon2010}, we aim to investigate what mechanisms are playing a role in powering a z$\approx$3.2 \lya\ nebulae at the center of a long-standing debate \citep[][hereafter ``LABn06"]{Nilsson2006}. The initial VLT/FORS data revealed revealed an isolated nebula that lacked any associated continuum or radio-loud counterparts, and the surface brightness profile of the nebula resembled those seen in the theoretical simulations of gravitationally cooling nebulae at the time. It was thus concluded that the nebular emission of this source was due to gravitational cooling. Subsequently, \cite{Prescott2015b} studied the environment of LABn06 using deep HST CANDELS imaging and 3D-HST grism spectroscopy as well as improved mid-infrared (MIR) photometry and photometric redshifts from {\it Spitzer} and {\it Herschel}. They identified 6 continuum sources associated with the nebula along with evidence for an obscured 
AGN in the vicinity, at the location of a nearby MIR 
source named ‘Source 6’ (hereafter S6). In this paper, we return to this controversial \lya\ nebula with newly obtained MUSE observations in order to investigate the morphology and kinematics of the nebula and what these findings tell us about the powering of the \lya\ emission.

In Section~\ref{sec:method}, we discuss our observations and data reduction, and in Section~\ref{sec:MappingEmission} we describe the 3D moment analysis of our data and the \lya\ line profiles across the nebula. In Section~\ref{sec:results}, we present our results on the morphology, kinematics, and emission line constraints. In Section~\ref{sec:discussion}, we discuss the nature of S6 and how this particular system compares with the larger population of \lya-emitting nebulae. We conclude in Section~\ref{sec:conclusion}. We assume the standard $\Lambda$CDM cosmology, i.e., $\Omega_m=0.27$, $\Omega_{\Lambda}=0.73$, $h=0.7$, correspondiing to an angular scale of 7.8 physical kiloparsecs/arcsec (pkpc/$^{\prime\prime}$) at z$\approx$3.2.

\section{Observations \& Data Reduction} \label{sec:method}
\subsection{MUSE Observations}
Spectroscopic observations of LABn06 were taken using the MUSE integral field spectrograph \citep{Bacon2010} on the Very Large Telescope (VLT) at the Cerro Paranal Observatory in Chile. The observations were carried out in 15 observing blocks (OBs) over 3 nights in 2019 (UT 2019 Jan 16, 28, \& 29) and 7 nights in 2021 (UT 2021 Feb 3, 4, 8, 9, 10, 14, \& 15). For the 2019 observations, each OB was split into two 1447 second exposures on the nebula, for a total of 4.02 hours of on-target exposure time. For the 2021 observations, each OB was split into two 1387 second exposures, for a total of 7.71 hours of on-target exposure time. In 2021, the pointing was shifted by $10$\arcsec\ to the West relative to that used in 2019 in order to ensure a sufficiently bright guide star for the Slow Guiding System (SGS). The combined 15 OBs resulted in a total of 11.73 hours of on-target exposure time. The pixel scale for MUSE observations is 0.2 arcsec/pixel, and the wavelength dispersion for each integral field unit (IFU) is 1.25\AA/pixel. Observations were taken using the nominal wavelength range (4800-9300\AA). The resolving power for the instrument ranges from R$\sim$1770 on the blue side to R$\sim$3590 on the red side. Details of these observations, including the seeing for each night, are listed in Table \ref{tab:observations}.
\begin{table}[]\centering
\begin{tabular}{lllll}
\hline
\hline
Date-Time         & AM    & DS       & $t_\mathrm{exp}$  & Sky\\
{[}yy/mm/dd-UT{]} &       & {[}''{]} & {[}s{]} &     \\ \hline
19/01/16-01:25:55 & 1.022 & 0.49     & 1447.0   &  THN \\
19/01/16-01:52:03 & 1.05  & 0.64     & 1447.0    &  THN\\
19/01/16-03:43:04 & 1.352 & 0.91     & 1447.0    &  THN\\
19/01/16-04:09:08 & 1.495 & 0.63     & 1447.0   &  THN \\
19/01/28-02:16:18 & 1.2   & 1.13     & 1447.0    &  THN\\
19/01/28-02:42:24 & 1.293 & 1.26     & 1447.0    &  THN\\
19/01/29-01:24:48 & 1.087 & 0.95     & 1447.0   &  THN \\
19/01/29-01:50:53 & 1.141 & 1.11     & 1447.0    &  THN\\
19/01/29-02:24:15 & 1.239 & 1.19     & 1447.0    &  THN\\
19/01/29-02:50:20 & 1.344 & 1.19     & 1447.0    &  THN\\ 
21/02/04-01:17:49 & 1.124 & 0.87     & 1387.0   &  THN \\
21/02/04-01:43:04 & 1.189 & 0.58     & 1387.0    &  THN\\
21/02/04-02:14:36 & 1.3 & 0.5     & 1387.0    &  THN\\
21/02/04-02:40:05 & 1.422 & 0.48     & 1387.0    &  THN\\
21/02/05-01:00:52 & 1.097 & 0.51     & 1387.0    &    CLR\\
21/02/05-01:26:08 & 1.153 & 0.34     & 1387.0    &    CLR\\
21/02/09-00:48:25 & 1.103 & 0.68     & 1387.0    &  THN\\
21/02/09-01:13:34 & 1.161 & 0.54     & 1387.0    &  THN\\
21/02/10-01:51:09 & 1.3 & 0.79     & 1387.0    &    CLR\\
21/02/10-02:16:17 & 1.421 & 0.87     & 1387.0    &    CLR\\
21/02/11-00:47:06 & 1.117 & 0.64     & 1387.0    &    CLR\\
21/02/11-01:12:33 & 1.18 & 0.59     & 1387.0    &    CLR\\
21/02/11-01:43:35 & 1.286 & 0.81     & 1387.0    &    CLR\\
21/02/11-02:08:42 & 1.401 & 0.92     & 1387.0    &    CLR\\
21/02/14-01:42:31 & 1.331 & 0.72     & 1387.0    &    CLR\\
21/02/14-02:07:42 & 1.462 & 0.57     & 1387.0    &    CLR\\
21/02/16-00:41:12 & 1.149 & 0.99     & 1387.0    &    CLR\\
21/02/16-01:06:21 & 1.222 & 0.8     & 1387.0    &    CLR\\
21/02/16-01:37:48 & 1.345 & 0.72     & 1387.0    &    CLR\\
21/02/16-02:03:16 & 1.482 & 0.74     & 1387.0    &    CLR\\
\hline
\hline
\end{tabular}
\caption{Observation details. AM = airmass, DS = Differential Image Motion Monitor (DIMM) Seeing measurement expressed as the full-width-half-maximum (FWHM) in arcseconds. Sky = Sky Transparency, where THN = thin clouds and CLR = clear conditions. The airmass and DIMM  seeing were taken from the image headers.}
\label{tab:observations}
\end{table}

\subsection{Data Reduction}
Observations of all 15 OBs were reduced using the MUSE ESOREX \citep[v2.6.2]{Freudling2013, Weilbacher2020} command line reduction pipeline. Each exposure was processed separately using the pipeline's standard calibration procedure. Initially, each exposure was bias-subtracted ($muse\_bias$), dark-subtracted ($muse\_dark$), flatfield-corrected ($muse\_flat$), and wavelength-calibrated ($muse\_wavecal$). The master twilight calibration was used to apply a 3D illumination correction to the data cube ($muse\_twilight$). All of these calibrations were applied to the science and standard star exposures. The recipe $muse\_scibasic$ was used to apply these basic calibrations to the science exposures. During this stage, the data were also astrometrically aligned to the MUSE WCS. For the astrometric calibration, a geometry table was used to compute the relative locations of each slice within the IFU. We used the geometry calibration file provided by the ESOREX reduction pipeline, as it has been found to show little variation over time \citep[][see Section 3.6]{Weilbacher2020}.

The recipe $muse\_standard$ was then used to create a standard star response file. The standard flux calibration was applied to each science exposure, and sky subtraction was performed in the $muse\_scipost$ recipe. The sky subtraction step requires a Line Spread Function (LSF) as input to create a model of the sky lines. Although the ESOREX pipeline provides a standard version of the LSF profile for the MUSE instrument, we chose to create our own in order to obtain the most accurate pipeline sky subtraction possible for our data. We computed the spatial offsets between all 30 exposures relative to a reference exposure using the recipe $muse\_exp\_align$, and we determined flux scaling factors for each exposure with respect to our best observation in order to remove the effects of variable observing conditions. We then combined the individual exposures into a final data cube, applying the spatial offsets and flux scaling factors, using the recipe $muse\_exp\_combine$. The exposures were weighted by their integration times for this combination resulting in approximately equal weights. We followed this initial reduction with a single round of self-calibration, which is necessary to remove any low-level instrumental signatures that may not have been removed during the standard calibration steps \citep{Bacon2017}. To do this, we used a white light image of the final MUSE datacube to create a 2D source mask, where only the brightest sources in the field-of-view (FOV) were masked. This 2D mask was then used as an input to perform self-calibration on each individual exposure, and the self-calibrated exposures were then recombined into a final datacube.

The MUSE reduction pipeline sky subtraction step (performed in $muse\_scipost$) in version 2.6.2 of the pipeline tends to leave sky residuals in the final datacube. Therefore, we used the ESO recommended post-processing tool Zurich Atmosphere Purge \citep[ZAP;][v.2.0]{Soto2016} to remove any pipeline residuals. We found that optimal sky subtraction was achieved using the ZAP {\it process} function with a median level sky subtraction method (zlevel=`median') and a continuum filter method for the singular value decomposition (SVD) computation (cftype = `median'; cfwidthSVD=200). To confirm that we were not over/under-subtracting the sky emission, we used an emission line source at the edge of our FOV for which the [OIII]4959/5007 doublet at 7965.25\AA\ happens to coincide with an optical skyline\footnote{\url{https://www.eso.org/observing/dfo/quality/UVES/uvessky/sky_8600L_6.html}}. We selected the number of eigenspectra that preserved the expected doublet line ratio for the [OIII] doublet \citep[f$_{5007}$/f$_{4959}$=3;][]{Storey2000}. We note, however, that for the \lya-emitting nebula at z$\approx$3.2 being investigated here, the \lya\ line is on the blue side of the spectrum where concerns about over-subtracted sky lines are minimal. 

\subsection{Flux Calibration Validation}\label{sec:bootstrap}

As a check on the flux calibration of our data, we compared our MUSE observations to GOODS-S/CANDELS \citep{Giavalisco2004, Koekemoer2011} F775W imaging mosaics produced by the 3D HST Team \citep{Brammer2012, Skelton2014, Momcheva2016}. We applied the HST F775W filter transmission curve (including instrumental response) to the MUSE data cube to create a pseudo F775W broadband image. To determine the point-spread-function (PSF) of the MUSE data, we fit two 1D Gaussian models in x and y to the brightest source in the FOV of the filtered MUSE 2D image (a bright, unresolved continuum source located at 03:32:12.644 -27:43:30.11) and measured the full-width-half-maximum (FWHM) of the MUSE PSF to be x=0.89\arcsec and y=0.89\arcsec. These measurements are consistent with the night-to-night seeing in Table~\ref{tab:observations}, which makes sense given that the source is compact. We convolved the HST mosaic with a PSF kernel matched to the MUSE PSF, and we convolved the MUSE image with a PSF kernel matched to the HST PSF, assuming the HST F775W PSF (0.105\arcsec) derived by \cite{Skelton2014}. Finally, we extracted aperture photometry of the source within a 3\arcsec\ radial aperture in both the HST and the MUSE data. The ratio between them provided a flux scaling of f$_{HST}$/f$_{MUSE}$=$0.98\pm 0.04$. We also performed this measurement for the second brightest source in the FOV (an emission line source located at 03:32:13.198 -27:42:40.57) resulting in a flux scale of f$_{HST}$/f$_{MUSE}$=$1.0\pm 0.07$. The weighted average of these two resulting ratios was f$_{HST}$/f$_{MUSE}$=$0.99\pm 0.03$, indicating accurate flux calibration for our final datacube.

\subsection{Variance Cube Correction}
A known issue in the ESOREX MUSE pipeline reduction is that it underestimates the variance cube when resampling the data values from PIXTABLE format to datacube format. This process introduces correlated errors that are not accounted for by the pipeline. \cite{Bacon2017} performed an experiment where they created a test PIXTABLE filled with perfect Gaussian noise (centered on 0.0 with a variance of 1.0), and then resampled this into a final datacube. This resulted in a pixel-to-pixel standard deviation of 0.6, implying the pipeline variance data needed to be multiplied by a correction factor of $(\frac{1}{0.6})^2$ in order to recover the true input variance. This correction factor was obtained using a pixfrac drizzle parameter of 0.8, the default value in the ESOREX MUSE pipeline and the value used for the reduction of our data. Therefore, we applied this correction factor to the variance extension of our final sky-subtracted datacube.

\subsection{Creating an Emission Line Cube} \label{sec:creatingEmLineCube}
To detect emission line sources in the datacube, we used the Line Source Detection and Cataloguing tool \citep[LSDcat;][]{Herenz2017}. The tool performs continuum source subtraction and then smooths the data both spatially (1.19’’) and spectrally (300 km/s). The smoothing kernel widths used for this step were chosen to match the maximum seeing during the observations and the spectral extent of the target \lya\ emission line, derived using visual inspection of the datacube. The maximum extent of the final datacube is 81.58\arcsec$\times$ 61.19\arcsec$\times$4601.25\AA; the portion with full exposure time depth spans dimensions of 60.76\arcsec$\times$ 61.19\arcsec$\times$4400.0\AA.

\begin{figure*}[htb!]
\centering
\subfloat[]{\includegraphics[width=3.5in]{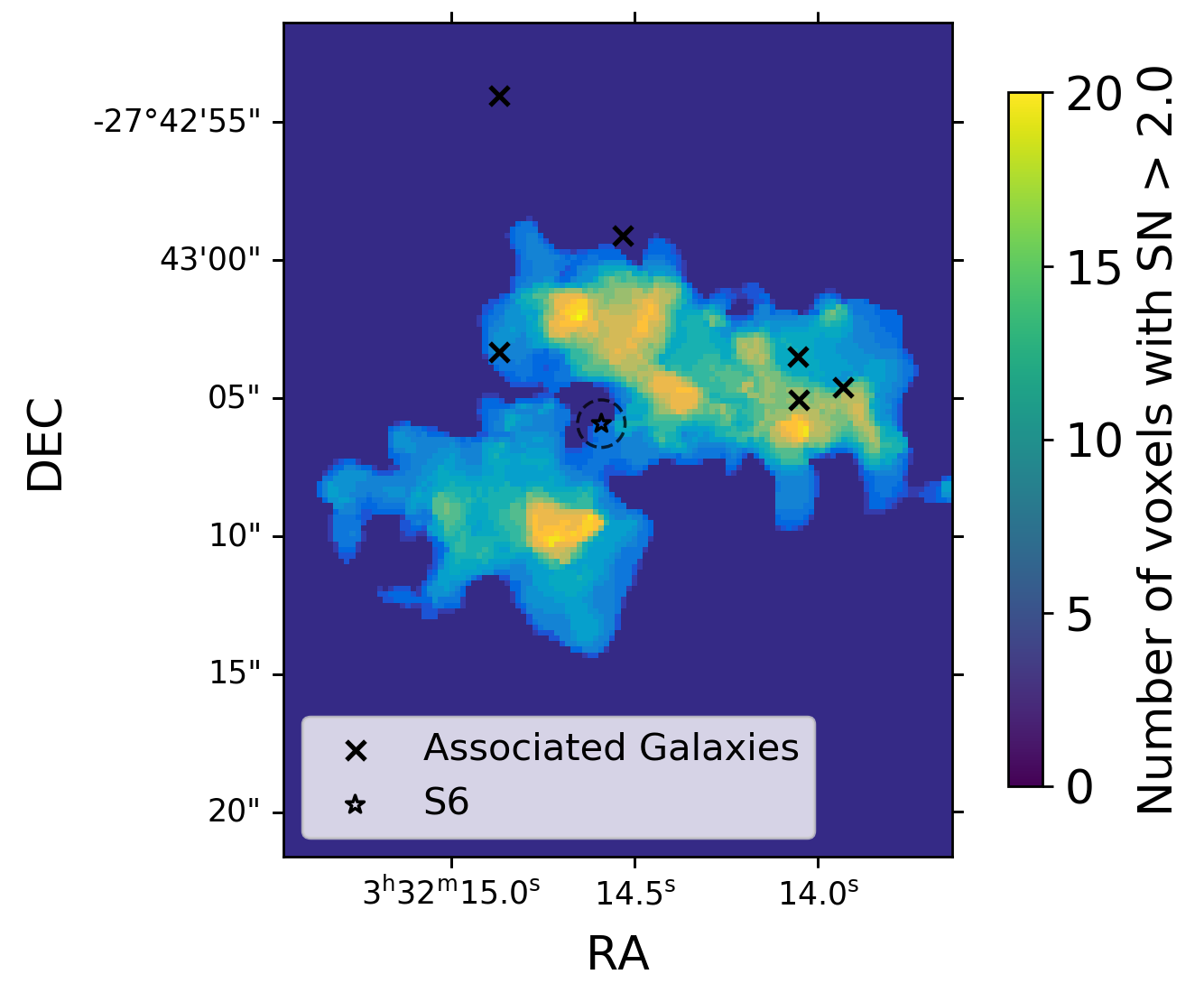}} 
\subfloat[]{\includegraphics[width=3.5in]{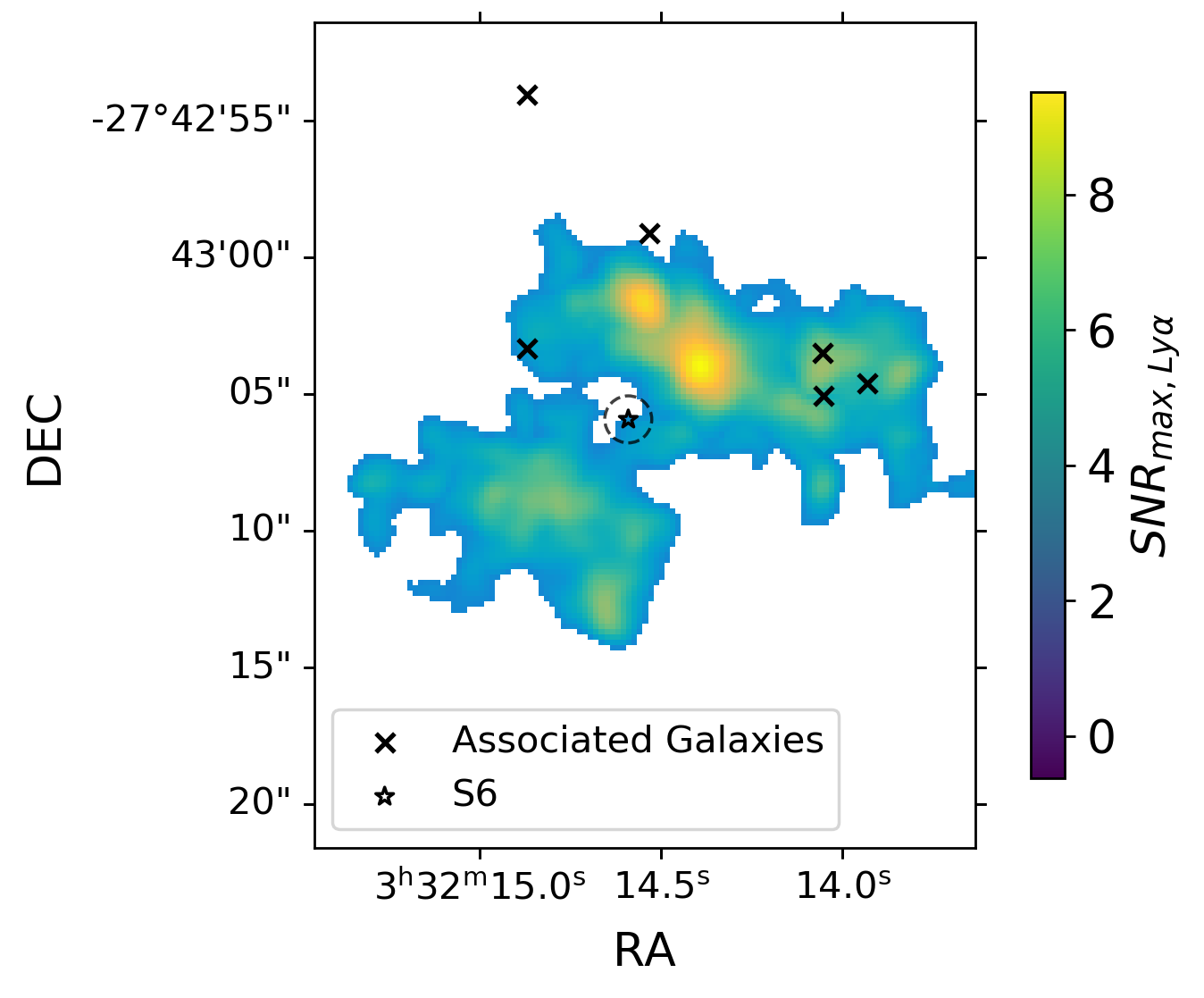}}
\caption{(a) 3D \lya\  mask collapsed along the wavelength direction showing the number of voxels with SNR $\geq$ 2 within a given spaxel. (b) Maximum SNR in the detected \lya\ emission along a given spaxel within the 3D mask. In each panel, we plot the locations of the 6 associated galaxies (black crosses) and S6 (black open star) identified by \cite{Prescott2015b}. The dashed circle represents the FWHM (1.72\arcsec) of the 4.6 $\mu$m Spitzer/IRAC data used to identify S6.}
\label{fig:em_significance} 
\end{figure*}

\subsection{Masking}
\label{sec:masking}
In order to construct a 3D mask for the analysis of this nebula, we followed the prescription detailed in \cite{Battaia2019}. Specifically, for each wavelength slice in a signal-to-noise ratio (SNR) cube, we identified voxels with SNR$\geq$2. The wavelength slice with the largest connected region of SNR$\geq$2 voxels was 5048.48\AA, which we therefore used to define the systemic velocity of the \lya\ emission ($z_{Ly\alpha}=3.15$). After identifying the wavelength slice containing the largest connected area, we stepped through the cube in the direction of both increasing and decreasing wavelength, attaching new segments to the 3D mask only if there was at least one SNR$\geq$2 voxel that overlapped spatially with the 3D mask defined in previous layers. In doing this, we obtained a final 3D mask composed of 30,983 voxels, extending across 31.25\AA, and centered on 5048.48\AA. The full extent of the connected SNR$\geq$2 region, measured as the maximum projected edge-to-edge distance of the collapsed mask, is 172.6 pkpc.

We display a voxel density map in Figure \ref{fig:em_significance} panel (a), showing the number of voxels along a given spaxel with SNR$\geq$2. In Figure \ref{fig:em_significance} panel (b), we show the maximum SNR along each spaxel. The locations of the 6 associated continuum sources and S6 are indicated \citep[][]{Prescott2015b}.

\section{Mapping the \lya\ Emission}
\label{sec:MappingEmission}

\subsection{Adaptive Narrowband \lya\ Image}

\begin{figure}
    \centering
    \includegraphics[width=\linewidth]{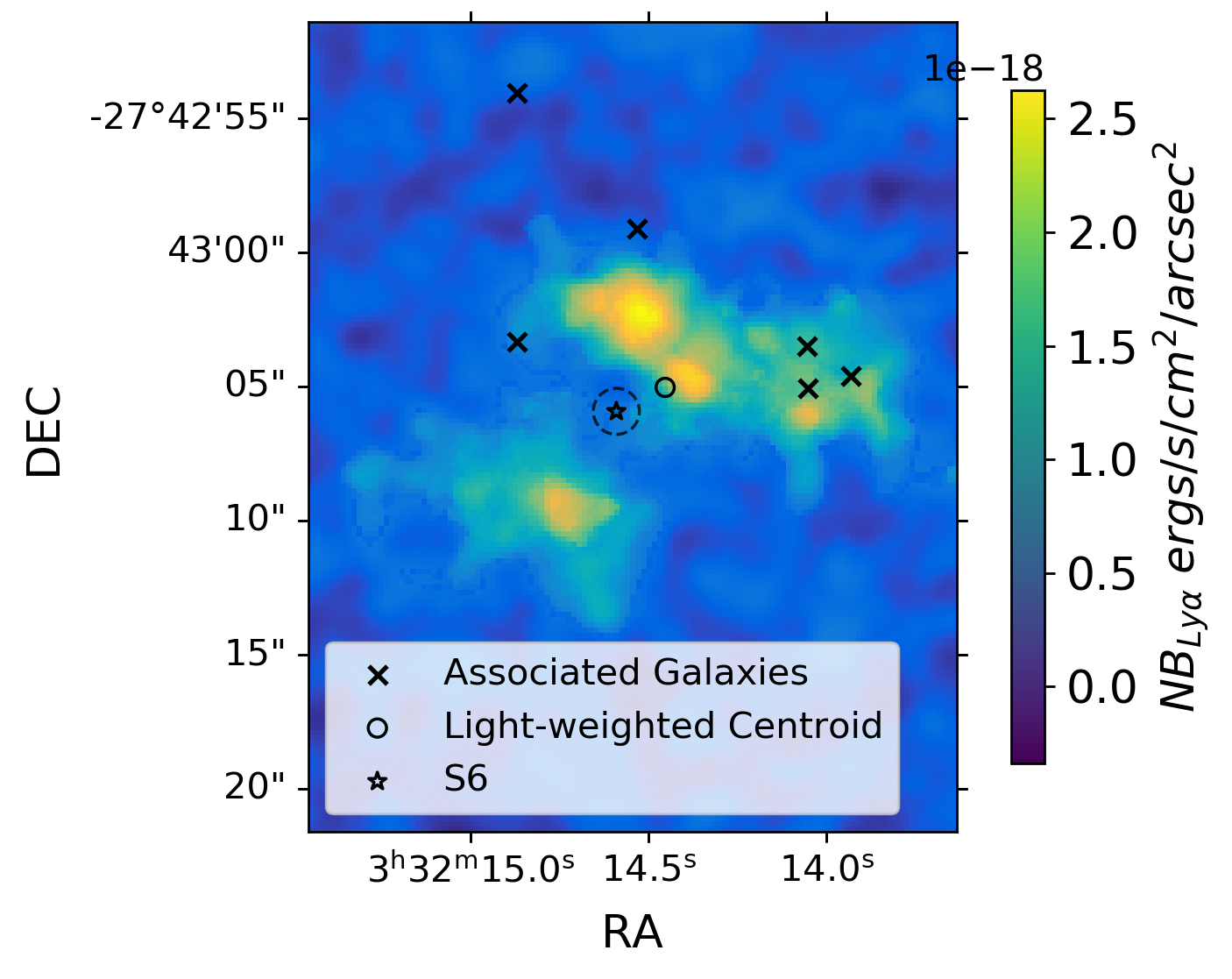}
    \caption{The adaptive narrowband map. Spaxels with SNR$\geq$2 are summed over all wavelengths contained in the 3D mask; spaxels with SNR less than the cutoff are summed over 5\AA\ centered on the central wavelength (5048.48\AA). We include the positions of the 6 associated galaxies and S6 identified by \citet[][]{Prescott2015b}, along with the light-weighted centroid of the \lya\ emission calculated from the adaptive narrowband map (open circle). The dashed circle represents the FWHM (1.72\arcsec) of the 4.6 $\mu$m Spitzer/IRAC data used to identify S6.}
    \label{fig:NBSB}
\end{figure}

To visualize the \lya\ nebula, we created an adaptive narrowband image following \cite{Herenz2020}. Voxels with SNR$\geq$2 are summed within a given spaxel, and spaxels that contain no voxels above this SNR threshold are summed along 5\AA\ centered on the central wavelength (5048.48\AA). The resulting adaptive narrowband image is shown in Figure \ref{fig:NBSB}. We note that this optimally-extracted narrowband image enhances the SNR of the object in a given spaxel because the filter width over which voxels are summed varies from spaxel to spaxel. This differs from standard narrowband imaging in which the filter width is fixed for all spatial positions across a source \citep[see][Appendix A, for a comparison]{Borisova2016}. We use this adaptive narrowband map to calculate the light-weighted centroid of the \lya\ emission and indicate this on all maps presented in this paper. We note that the light-weighted centroid is unchanged if we instead use the MUSE datacube to create a standard narrowband image of the \lya\ emission with a fixed width of 31.25\AA. 

\subsection{Moment Analysis and Line Profile Shape}
\label{sec:momentprofile}
\begin{figure*}[htb!]
\centering
\subfloat[]{\includegraphics[width=3.1in]{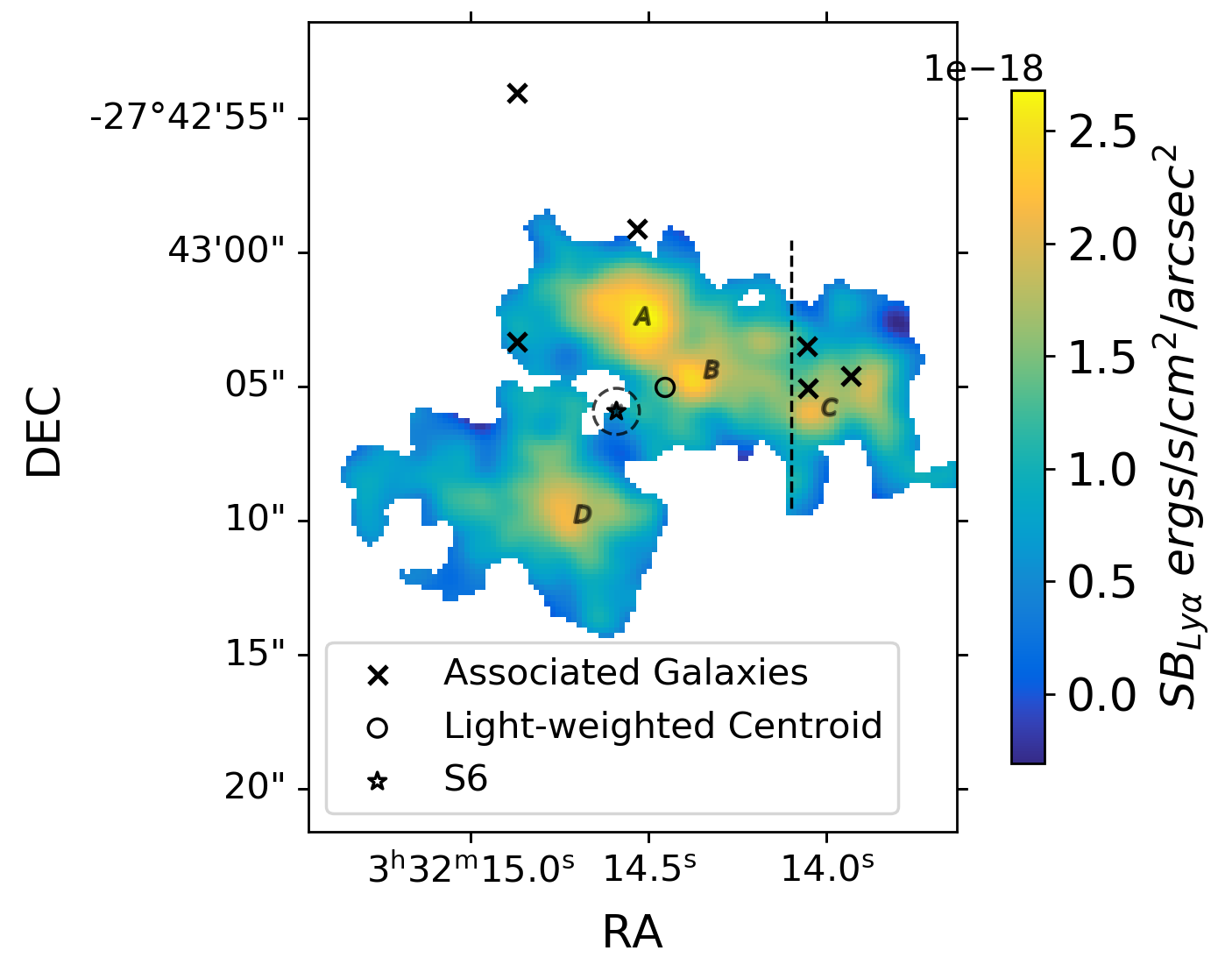}} 
\subfloat[]{\includegraphics[width=3.1in]{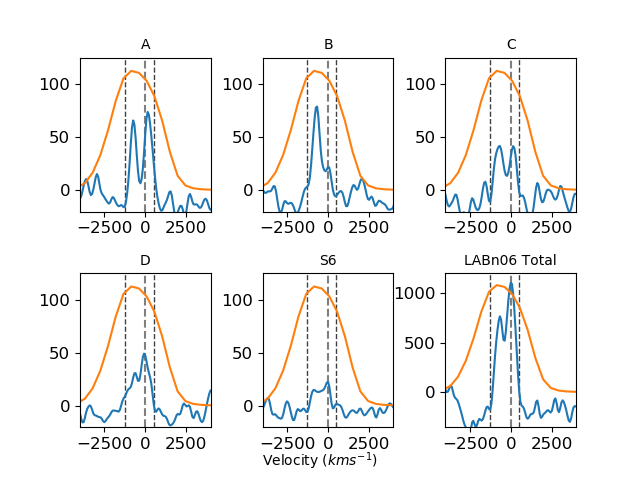}}
\caption{Regions of interest within the nebula (A-D and the location of S6). (a) 0th moment map with regions of interest labeled. The vertical dashed line marks a dip in \lya\ surface brightness between region B and the western lobe as discussed in Section~\ref{sec:Results1}. (b) Spectra extracted from the labeled regions in panel (a) using a diameter aperture of 12 native MUSE pixels (2.4 arcsec) and from the entire nebula. The gray long-dashed vertical line marks the systemic velocity defined as the central wavelength slice (5048.48\AA), and the two black short-dashed vertical lines mark the widest wavelength range of the 3D \lya\ mask. Each spectrum is plotted over 8000 km s$^{-1}$ and over arbitrary flux units on the y-axis. The transmission curve for the narrowband filter used in the discovery paper by \cite{Nilsson2006} is over-plotted (orange). The \lya\ line profiles across the nebula are varied and complex, showing multiple peaks or asymmetric line shapes.} 
\label{fig:spec_samples} 
\end{figure*}

We calculated the 0th, 1st, and 2nd moments of our data in order to investigate the surface brightness profile and kinematics of the emitting nebula. Only voxels within the 3D mask with SNR$\geq$2 are included in these moment calculations. The 0th moment is shown in Figure~\ref{fig:spec_samples} panel (a); this is identical to Figure~\ref{fig:NBSB}, but contains only spaxels within the 3D mask. Particular regions of interest, selected as the brightest knots of \lya\ emission across the nebula, are indicated by markers A-D, along with the location of S6. In Figure~\ref{fig:spec_samples} panel (b) we show the \lya\ line profiles for each region of interest, measured within an aperture diameter of 12 native MUSE pixels (2.4 arcsec).

\begin{figure*}
\centering
\subfloat[]{\includegraphics[width=3.1in]{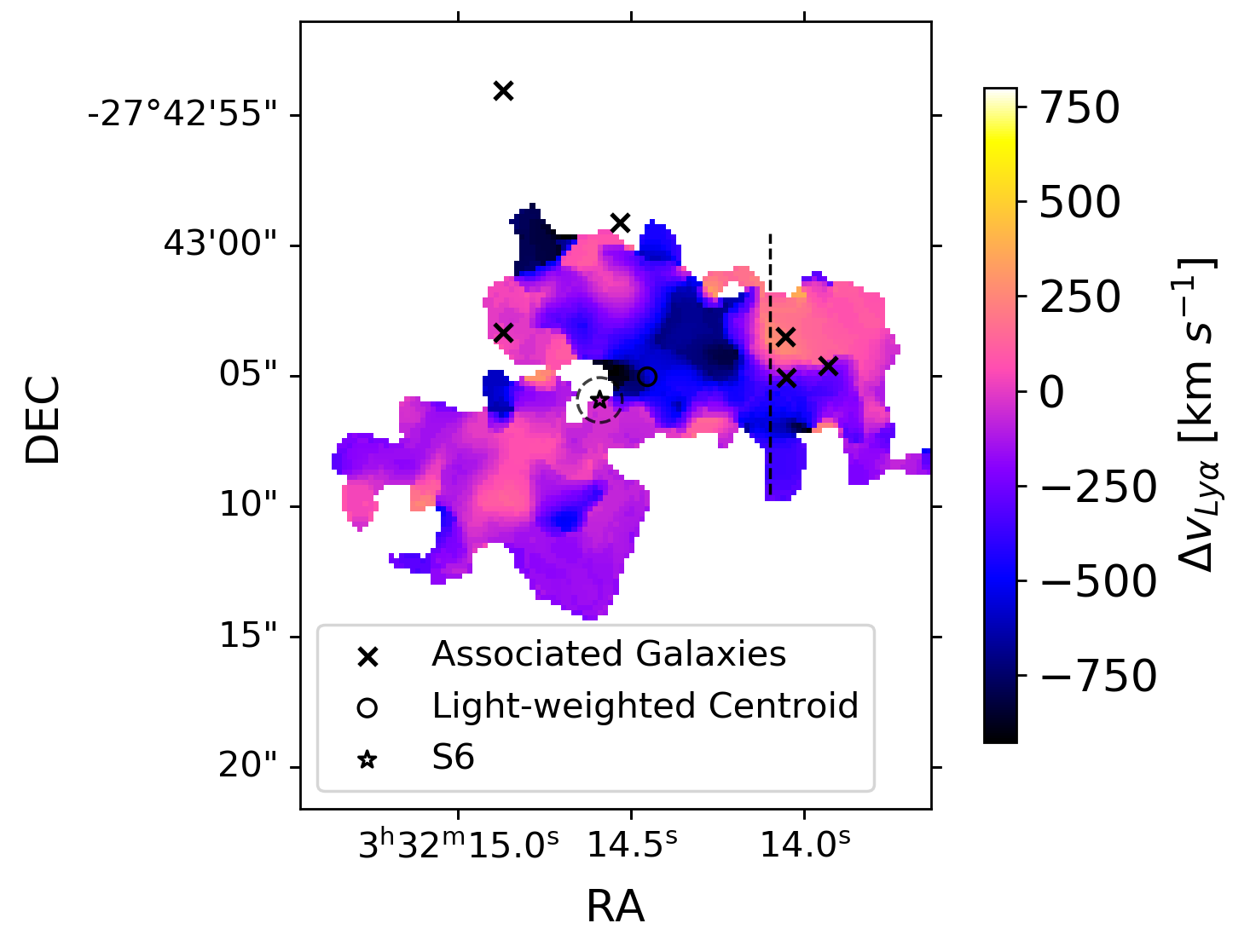}} 
\subfloat[]{\includegraphics[width=3.1in]{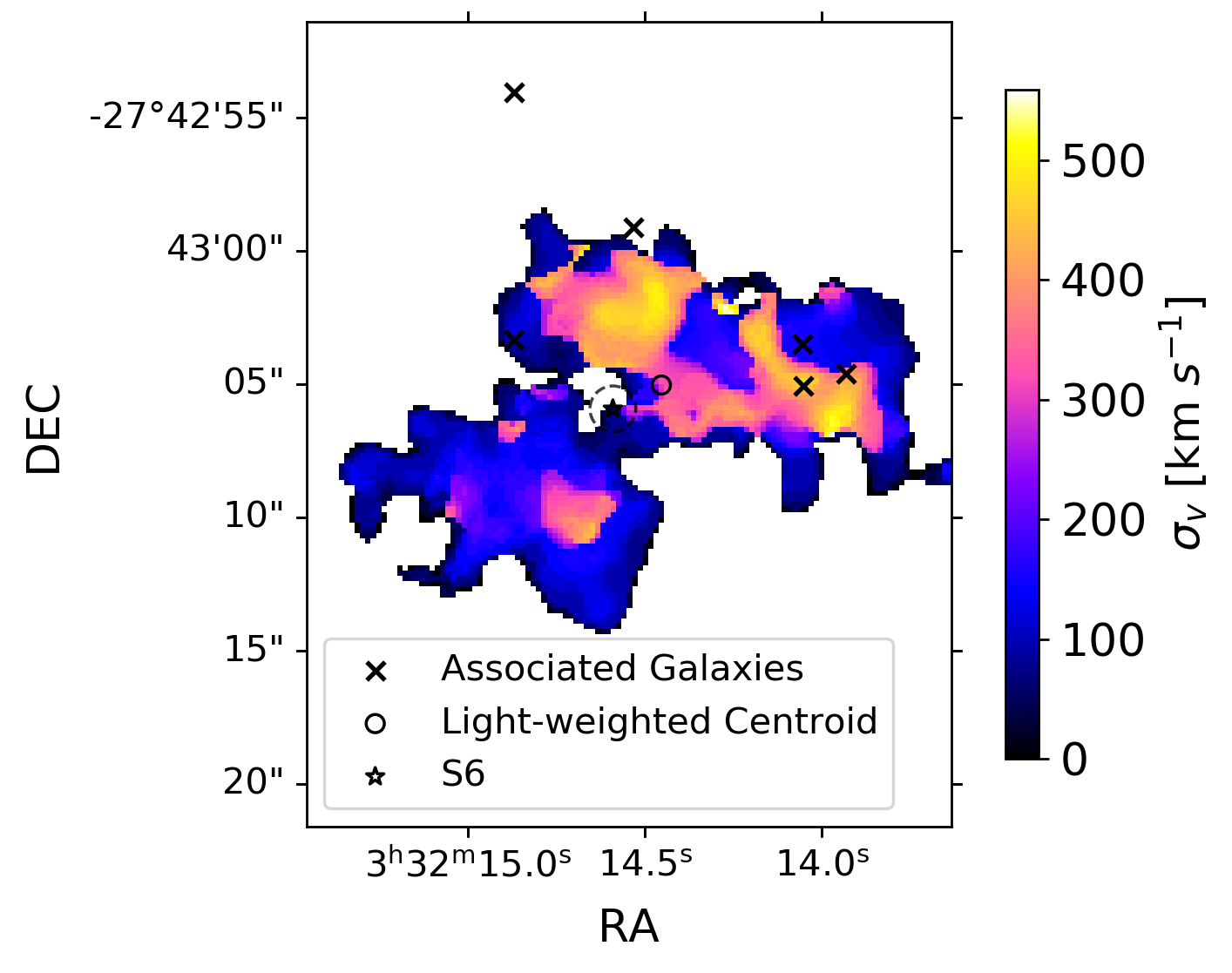}}
\caption{First and second image moment maps. In each case, the systemic velocity 
was defined to be the wavelength layer with the largest connected region of voxels with SNR$\geq2$ (5048.48\AA). (a) The velocity map derived from the first moment of the flux data cube. The kinematics of the gas in LABn06 show hints of ordered motion around the location of S6. There is also a steep North-South velocity gradient seen across the western lobe (see Section~\ref{sec:Results1}) of the nebula near three of the associated galaxies \citep[][]{Prescott2015b}. The vertical dashed line marks a dip in \lya\ surface brightness between region B and the western lobe as discussed in Section~\ref{sec:Results1}. (b) The apparent velocity dispersion map ($\sigma_{V}$) derived from the second moment of the \lya\ emission. The gas to the South-East of S6 shows lower velocity dispersions. Larger velocity dispersions are seen to the North-West of S6, but these higher $\sigma_v$ values are likely caused by the double-peaked nature of the \lya\ emission in this region.}
\label{fig:kinematics} 
\end{figure*}

The 1st and 2nd moments, 
corresponding to the velocity centroid and standard deviation, are shown in Figure~\ref{fig:kinematics}, panels (a) and (b), respectively. Many previous analyses of \lya-emitting halos have chosen to display a FWHM map as a measure of the line width. This is done by assuming a Gaussian emission line profile, measuring the standard deviation from line center, and multiplying this by $\approx$2.355, the conversion factor between standard deviation and FWHM for a Gaussian. As seen in Figure~\ref{fig:spec_samples}, the assumption of a Gaussian profile is not necessarily valid for \lya\ emission lines; the profiles often show multiple peaks or enhanced tails, likely due to the resonant nature of \lya. Thus, we instead chose to show the standard deviation directly as an ``apparent velocity dispersion'' map, without converting to FWHM or making any assumptions about line shape \citep[][]{Herenz2020}.

\section{Results} \label{sec:results}
\subsection{Morphology and Size}\label{sec:Results1}

From the maps presented in Figure~\ref{fig:em_significance} and \ref{fig:NBSB}, we can see multiple regions of significant \lya\ emission, but the brightest parts of the nebula are not coincident with the associated galaxies in the region.  
Instead, LABn06 consists of a two-lobed structure centered on the position of S6, which lies within 2.2\arcsec (17.2 pkpc) of the light-weighted centroid of the nebula. Assuming the source powering the emission lies near the light-weighted centroid of the nebula, the two lobes could be tracing out a bipolar structure with a wide opening angle.

We quantify the diameter of the two-lobe structure using the distance from region D to region A (as labeled in Figure~\ref{fig:spec_samples}) -- a distance of 7.8\arcsec\ (61 pkpc). Some low signal-to-noise emission appears to extend over to the West of the light-weighted centroid, connecting the brightest Northern peaks (A and B) to the three associated galaxies near region C. The vertical dashed line in Figure~\ref{fig:spec_samples} marks a dip in the \lya\ surface brightness distribution. We will refer to the material west of this vertical dashed line as the ``western lobe'' throughout the remainder of the paper. The maximum source size down to the limit of our data, as defined in Section~\ref{sec:masking}, is 172.6 pkpc.

\subsubsection{Surface Brightness Profile}
\begin{figure*}
\centering
\subfloat[]{\includegraphics[width=3.1in]{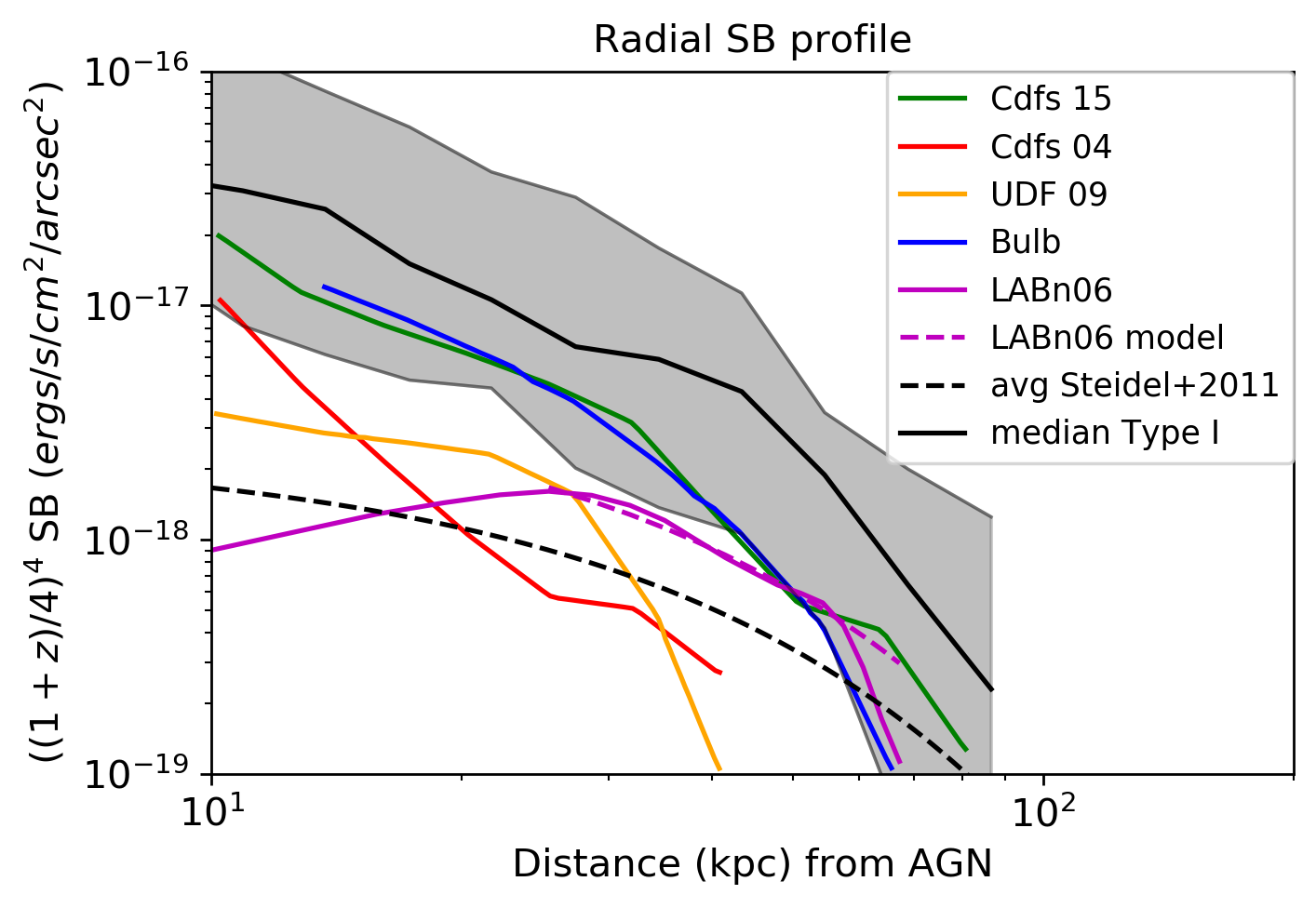}} 
\subfloat[]{\includegraphics[width=3.1in]{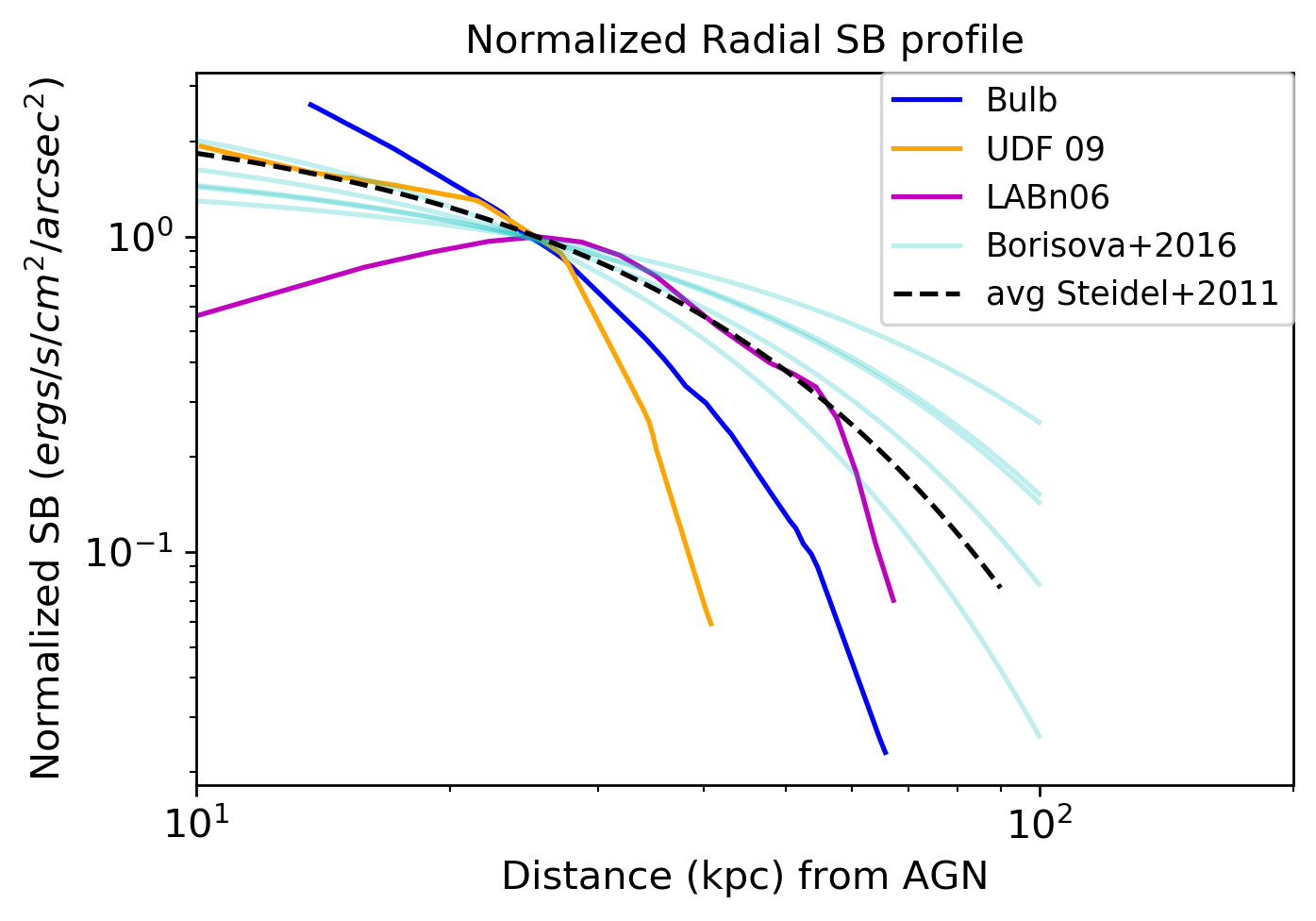}}
\caption{Circularly averaged surface brightness profiles all corrected to z=3. We only show profiles for distances greater than 10 pkpc (1.2\arcsec) as smaller radii are compromised by the seeing. (a) Comparisons between LABn06 (pink) and other Type II AGN observed with MUSE \citep[green, red, yellow, and blue solid lines;][]{denBrok2020}. The profile for LABn06 is best fit by an exponential profile with a scale radius of 24.2 pkpc (dashed pink line). We also plot the median of a sample of Type I radio-quiet and radio-loud quasars \citep[][black solid line]{Borisova2016}; the 10th and 90th percentiles of the Type I AGN distribution is given by the gray shaded region. We also show the \lya\ surface brightness profile derived using a stack of Lyman Break Galaxies \citep[][dashed black]{Steidel2011}. (b) Surface brightness profiles for the subset of sources well-fit by exponential profiles, normalized to 25 pkpc.}
\label{fig:SBprofiles} 
\end{figure*}

It is interesting to investigate how the surface brightness profile of LABn06 compares to other \lya-emitting sources in the literature. In Figure~\ref{fig:SBprofiles} panel (a), we show the circularly-averaged, radial surface brightness profile for the nebula centered on the location of S6. For comparison, we also show the measured profiles for a stacked sample of Lyman Break Galaxies \citep[][]{Steidel2011}, a sample of \lya\ halos surrounding 2 radio-loud and 16 radio-quiet Type I quasars \citep[][]{Borisova2016}\, and four halos around Type II AGN \citep{denBrok2020}. All profiles displayed are redshift-dimming-corrected to z=3 by the factor $(\frac{1+z}{4})^4$. 

The most striking finding is that the inner region of LABn06 appears distinct from all of the other profiles, showing a dip in the surface brightness at the location of S6. Outside of 25 pkpc, the profile resembles that of other nebulae. To make a quantitative comparison, we fit both a power law and an exponential profile to LABn06, only including surface brightness values at radial distances greater than 25 pkpc (the distance of peak A) from S6. The nebula is best fit by an exponential with a scale radius of 24.2 pkpc ($\chi^2=$6.5). We compare LABn06 to a subset of sources well fit by exponential profiles in Figure~\ref{fig:SBprofiles} panel (b), which emphasizes the distinct dip in the center of LABn06 compared to other sources. 

\subsubsection{Asymmetry}
From Figures~\ref{fig:em_significance}, \ref{fig:NBSB}, and \ref{fig:spec_samples} panel (a), it is obvious that LABn06 is not completely circularly symmetric, instead showing a bipolar structure. In an attempt to quantify the level of asymmetry, we apply both a moment-based method and a Fourier Decomposition method, as detailed in \cite{denBrok2020}. 

For the moment-based method, we calculate the second-order moment of the spatial pixels with respect to the known location of S6, including all pixels contained by the collapsed 3D mask. We do not include flux-weighting of the pixels in order to emphasize how diffuse emission is affecting the asymmetry. These second-order moments are then used to quantify the circular asymmetry of the nebula via the dimensionless parameter $\alpha$ \citep[][]{denBrok2020}:

\begin{equation}
    \alpha = \frac{1-\sqrt{(M_{xx}-M_{yy}^2)+(2M_{xy})^2}}{1+\sqrt{(M_{xx}-M_{yy}^2)+(2M_{xy})^2}}
\end{equation}

where M$_{xx}$, M$_{yy}$, and M$_{xy}$ are the second-order image moments. A value of $\alpha=1$ corresponds to a circularly symmetric nebula while $\alpha<1$ corresponds to a less circularly symmetric distribution of emission. For LABn06, we find $\alpha=0.60$, indicating a degree of asymmetry comparable to other Type II AGN ($\alpha=$0.4-0.7) but outside the 25 - 75 percentile range for Type I AGN ($\alpha=$0.65-0.85) \citep[][]{denBrok2020}.

The Fourier Decomposition method quantifies the nebula asymmetry as a function of distance from the known location of S6. Following  \citet[][]{denBrok2020}, we first reproject the (x,y) pairs of the surface brightness pixels onto a grid of (r,$\theta$) values. We then decompose $SB(r, \theta)$ into its Fourier series and solve for the Fourier coefficients $a_k(r)$ and $b_k(r)$. These coefficients can then be combined into a single coefficient: 
\begin{equation}
c_k(r) = \sqrt{a_k(r)^2+b_k(r)^2}  
\end{equation}

A circularly symmetric distribution will be dominated by the 0th order coefficients ($a_0$, $b_0$). 
By contrast, an elliptical profile centered on S6 will have a significant contribution from 2nd order coefficients ($a_2$, $b_2$), and an off-center elliptical profile will have a significant contribution from the 1st order coefficients, ($a_1$, $b_1$). Therefore, the ratio $\frac{c_k(r)}{c_0(r)}$ quantifies how much the profile diverges from a circularly symmetric one at a given radial distance. 

\begin{figure*}
    \centering
    \includegraphics[width=\linewidth]{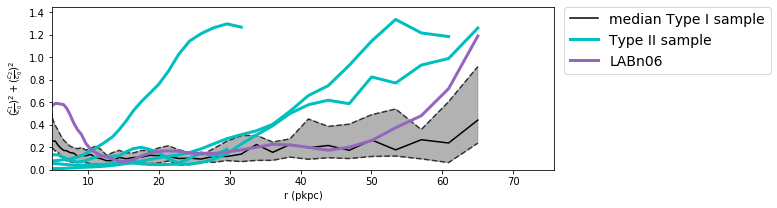}
    \caption{Fourier decomposition of the LABn06 surface brightness map as a function of distance from the location of S6 (purple). The dependent variable combines the first three fourier coefficients of the decomposed surface brightness distribution, where the 0th order coefficient corresponds to a circularly symmetric distribution and coefficients 1 and 2 quantify deviations from circular symmetry, as discussed in the text. For comparison, we plot the same profile for a sample of 4 Type II AGN (cyan) and a median profile for a sample of 19 Type I AGN \citep[][]{denBrok2020}. The dashed black lines bounding the gray shaded region indicate the 25th and 75th percentiles around the Type I median. LABn06 shows a hint of asymmetry at small radii ($r<10$ pkpc), and a rising asymmetry profile at large radii ($r>50$ pkpc), approaching what is seen in the Type II AGN sample.}
    \label{fig:FDecomp}
\end{figure*}

In Figure~\ref{fig:FDecomp}, we display $(\frac{c_1(r)}{c_0(r)})^2+(\frac{c_2(r)}{c_0(r)})^2$ for our nebula along with profiles of Type I and Type II AGN samples \citep[][private communication]{denBrok2020}. We find that the nebula shows higher asymmetry at very small radii ($r<10$ pkpc), although this region is somewhat compromised by the seeing limit of the data.  LABn06 is fairly symmetric at $r=10-50$ pkpc, but begins to deviate from circular symmetry at larger distances from S6 ($r>50$ pkpc).

\subsection{Kinematics}\label{sec:kinematics}
Overall, the velocity field (1st moment) and apparent velocity dispersion (2nd moment) maps in Figure~\ref{fig:kinematics} give the impression of relatively quiescent \lya-emitting gas in LABn06. Most of the velocity values lie in the range V$\approx$[-500,250] km s$^{-1}$, where 0 km s$^{-1}$ corresponds to $z_{Ly\alpha}=3.15$. A hint of ordered motion is apparent as a gradient between positive velocity values (V$\approx$250 km s$^{-1}$; pink) to the South-East of S6 and negative velocity values (V$\approx$-500 km s$^{-1}$; blue) to the North-West of S6. The one exception is the steep velocity gradient seen across the western lobe of the nebula, where the velocities range from -450 to 250 km~s$^{-1}$ over a distance of about $\sim$5\arcsec\ ($\sim$40 pkpc). The western lobe is in the vicinity of three of the associated galaxies, including the brightest one and the only one that is spectroscopically confirmed \citep{Prescott2015b}. This steep velocity gradient could be evidence that the gas in the western lobe is kinematically distinct from the gas centered on S6 and is actually linked with the 3 associated galaxies in the western lobe. 

The apparent velocity dispersion map shows a range of line widths from $\sigma_v\sim$200 km s$^{-1}$ to $\sigma_v\sim$500 km s$^{-1}$. The regions of highest apparent velocity dispersion seem to be concentrated mostly to the North of S6's MIR position as well as in the western lobe of the nebula. However, this is likely driven by the presence of double peaked line profiles in these regions (Section~\ref{sec:momentprofile}).

\subsection{Upper Limits on NV, CIV, HeII, \& CIII]}

\begin{figure*}
\centering
\subfloat[]{\includegraphics[width=0.2455\textwidth]{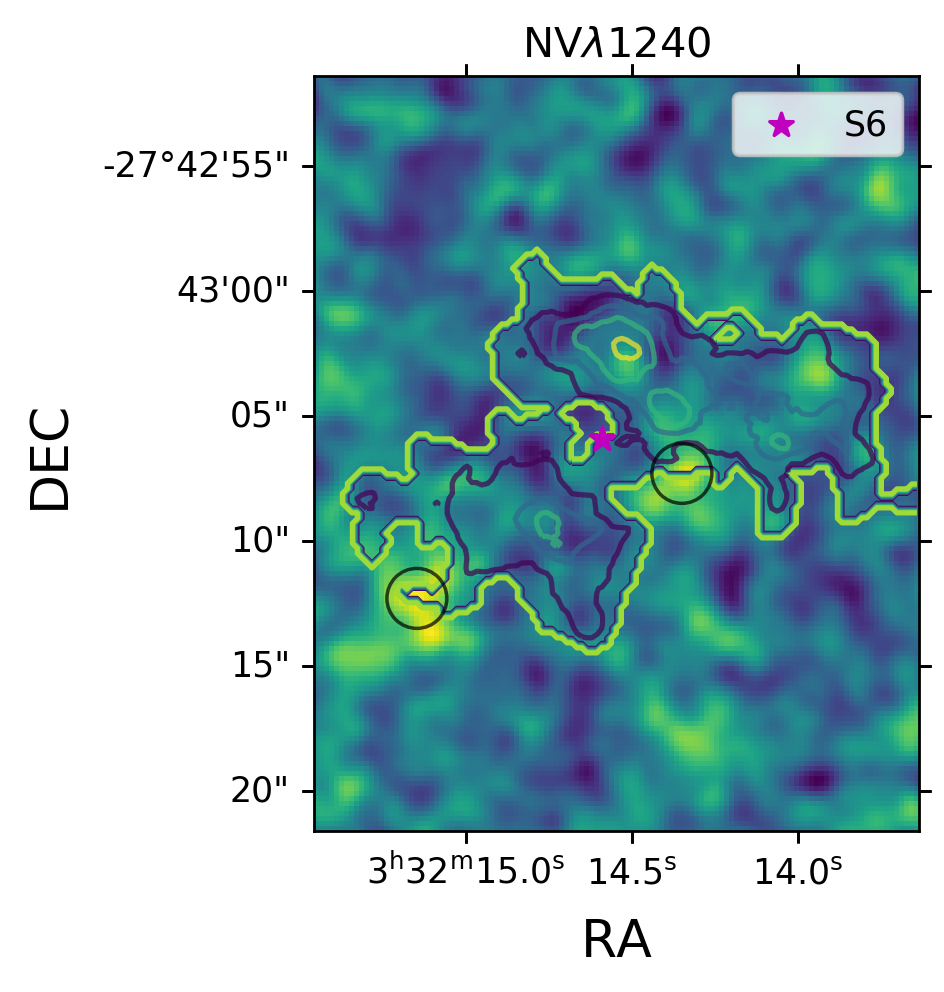}}
\subfloat[]{\includegraphics[width=0.227\textwidth]{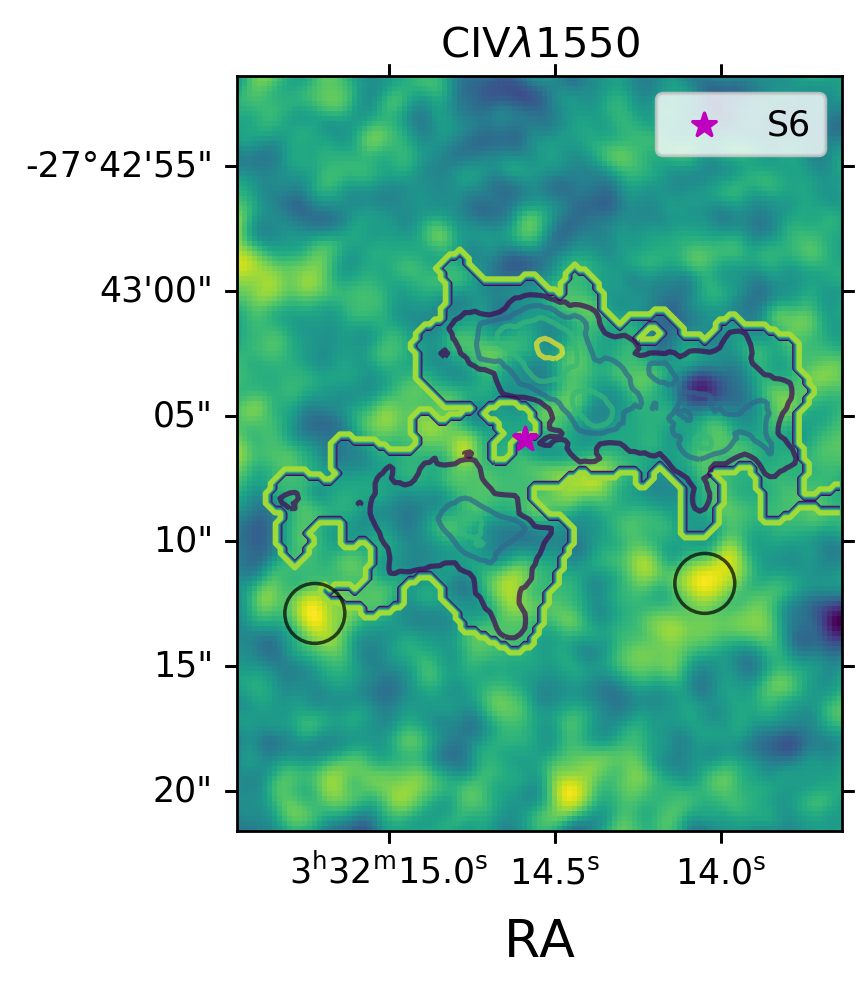}}
\subfloat[]{\includegraphics[width=0.227\textwidth]{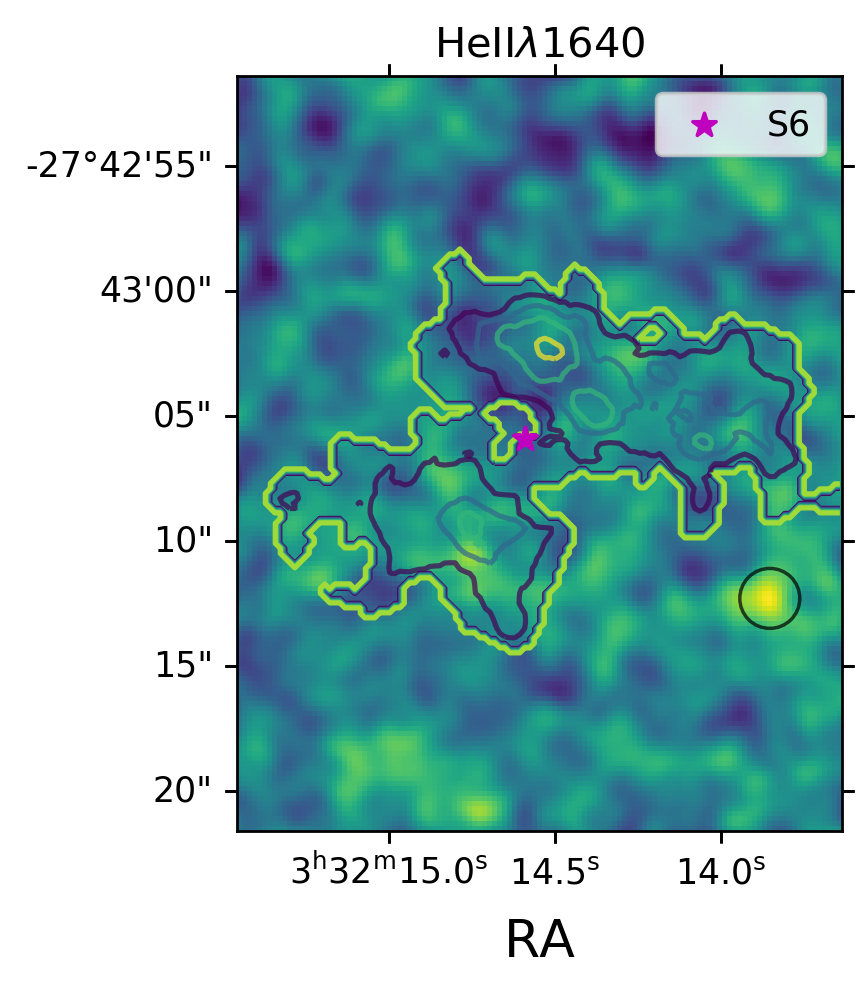}}
\subfloat[]{\includegraphics[width=0.297\textwidth]{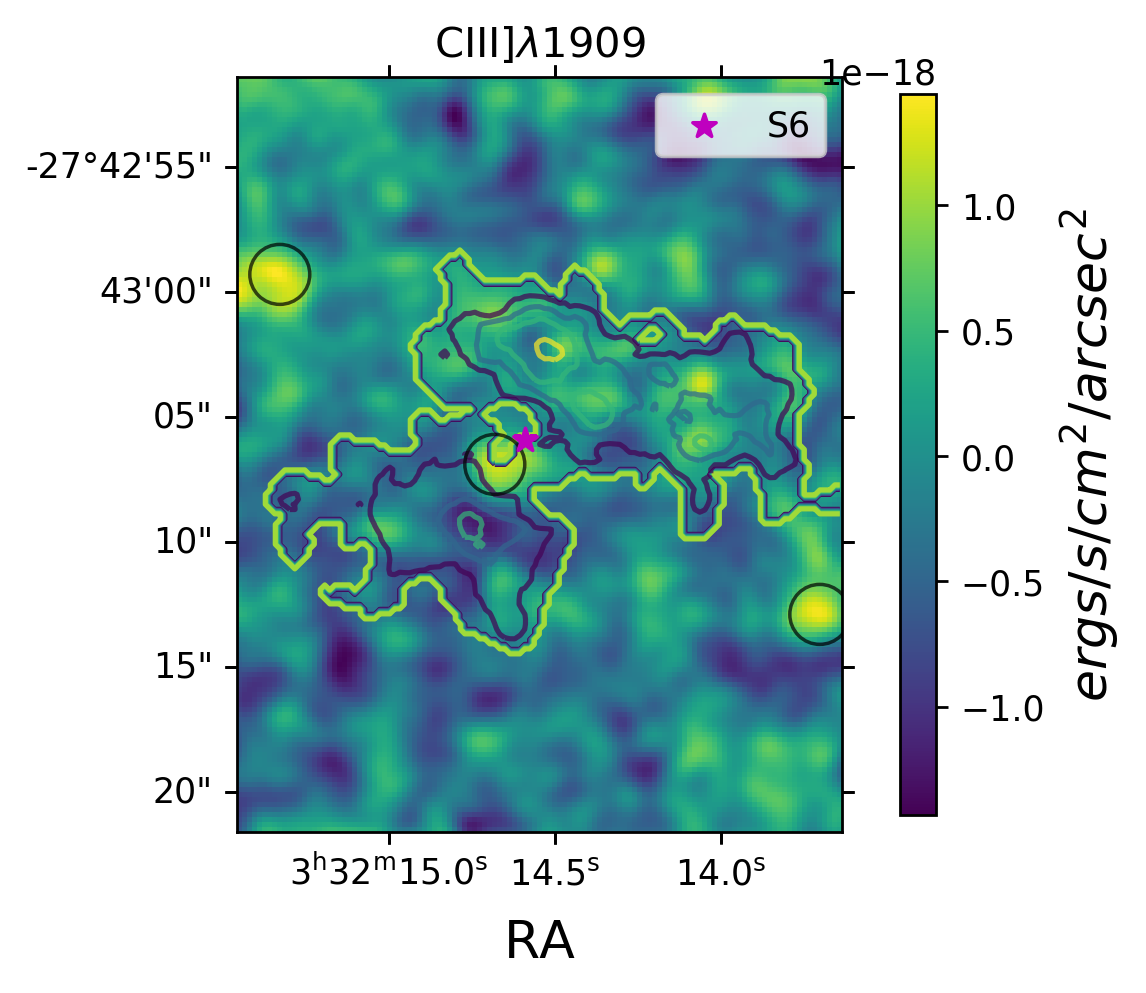}}
\caption{Flattened 2D images centered on the expected wavelength for each UV metal line assuming a source redshift of $z_{Ly\alpha}=3.15$. Contours of the \lya\ emission are plotted at arbitrary levels. We also indicate S6 (purple star) and the locations of some of the largest peaks in these images (black circles; 1.2\arcsec radius apertures). S6 seems to lie close to a peak in the NV$\lambda$1240 image and one of the peaks in the CIII]$\lambda$1909 image.}
\label{fig:flattened_metal_detections}
\end{figure*}

AGN are a common powering mechanism for \lya\ nebulae. To test the AGN-powering hypothesis in LABn06, we searched the data for additional rest-frame ultraviolet (UV) emission lines that would be indicative of an AGN \citep[i.e., NV$\lambda$1240, CIV$\lambda$1550, HeII$\lambda$1640, CIII\mbox{]}$\lambda$1909,][]{Feltre2016}. We inspected the datacube in regions contained by the 3D mask at the locations where one would expect to find these lines given a source redshift of $z_{Ly\alpha}=3.15$. After aligning the center of the 3D mask created in Section \ref{sec:masking} to the expected observed wavelength for each line, we summed over the wavelengths included in the 3D mask (a region of 31.25\AA\ across) to create a flattened 2D image. These images are displayed in Figure~\ref{fig:flattened_metal_detections} with several higher SNR peaks indicated by black circles. Peaks outside the lowest surface brightness contour of the nebula all appear to be linked with foreground and background HST sources not associated with this source. Interestingly, within the \lya\ nebula the closest source to the peaks in the NV$\lambda$1240 and CIII]$\lambda$1909 images is S6. We measure the flux of these peaks within a 1.2\arcsec (6 pixel) radial aperture centered on the peak and find $f_{NV}$= 3.05$\times10^{-18}$ erg s$^{-1}$ cm$^{-2}$ (SNR$=3.1$) and $f_{CIII]}$= 3.62$\times10^{-18}$ erg s$^{-1}$ cm$^{-2}$ (SNR$=2.4$). We estimated the uncertainty on these measurements by laying down 10000 random apertures across the images and taking the standard deviation of this distribution.

We also measured upper limits on all four of these lines following the scanning method of \citet[][Section 4.5]{Borisova2016} to allow for the possibility of kinematic offsets between \lya\ and other emission lines. For each line, we used the previously defined 31.25\AA-wide ($\approx$1180-1860 km s$^{-1}$) 3D mask (Section~\ref{sec:masking}) to scan a $\pm 3000$km s$^{-1}$ ($\approx$100-160\AA) wide region centered on the expected observed wavelength of the line assuming a source redshift of $z_{Ly\alpha}=3.15$. Beginning with the lower wavelength edge of the mask located -3000 km/s from the expected line center, we summed all of the voxels contained by the mask. We then shifted the entire mask by one spectral unit (1.25\AA), summed over the 3D mask, and repeated this procedure until the upper wavelength edge of the mask was +3000 km/s from the expected line center. Calculating the rms of these summed values provides us with a conservative measurement of the upper limit on the flux of these lines within the \lya\ nebula. These results for the full nebula are summarized in Table \ref{tab:tightupperlimits}. To make sure that using the entire 31.25\AA\ mask was not overly conservative, we repeated this scanning method with only the central 15\AA\ ($\approx$570-890 km s$^{-1}$) and 7.5\AA\ ($\approx$280-450 km s$^{-1}$) of the mask. In doing this, we found the same SNR peaks as indicated by black circles in Figure~\ref{fig:flattened_metal_detections}.

To allow for the possibility that any NV, CIV, HeII, or CIII] emission is confined only to the brightest regions of the \lya\ nebula, we also modified the previous method to incorporate an aperture-based approach. In repeating the scanning process described above, we focused on the specific regions of interest (A-D, S6) shown in Figure~\ref{fig:spec_samples}. That is, we summed over the 3D mask but only included voxels within a 1.2 arcsec radial aperture centered on the regions of interest. We then measured the rms of these summed values to derive an upper limit for the NV, CIV, HeII, and CIII] flux within each region A-D as well as the region centered on S6. These results are summarized in Table \ref{tab:tightupperlimits}.

\begin{deluxetable*}{cccccc}[htb!]
\centering
\tablewidth{0pt}
\tablehead{ \colhead{Region} & \colhead{\lya} & \colhead{$NV_{limit}$} & \colhead{$CIV_{limit}$} & \colhead{$HeII_{limit}$} & \colhead{$CIII]_{limit}$} \\}
\startdata
Full & 175.35 & $<$28.57 & $<$16.18 & $<$26.61 & $<$11.24 \\
A & 10.73 & $<$2.85 & $<$1.25 & $<$1.89 & $<$1.89 \\
B & 8.59 & $<$2.63 & $<$1.81 & $<$1.63 & $<$1.54 \\
C & 7.48 & $<$1.56 & $<$1.41 & $<$2.3 & $<$2.48 \\
D & 8.07 & $<$2.15 & $<$1.72 & $<$2.46 & $<$4.78 \\
S6 & 3.23 & $<$0.28 & $<$0.43 & $<$0.46 & $<$0.66 \\
\enddata

\caption{Flux measurements for \lya\ and 3$\sigma$ upper limits for NV, CIV, HeII, and CIII]. Measurements were taken from the smoothed flux data cube with the 3D \lya\ mask applied. Upper limit measurements were found by scanning the expected location of NV, CIV, HeII, and CIII]. The full nebula flux measurements were derived from an area of 165.36 arcsec$^2$. For regions A-D and the location of S6, the flux measurements were taken from 4.52 arcsec$^2$ (1.2\arcsec radius) circular apertures. All fluxes are in units of 10$^{-18}$erg s$^{-1}$ cm$^{-2}$.}
\label{tab:tightupperlimits}
\end{deluxetable*}

\begin{figure*}
    \centering
    \includegraphics[width=\linewidth]{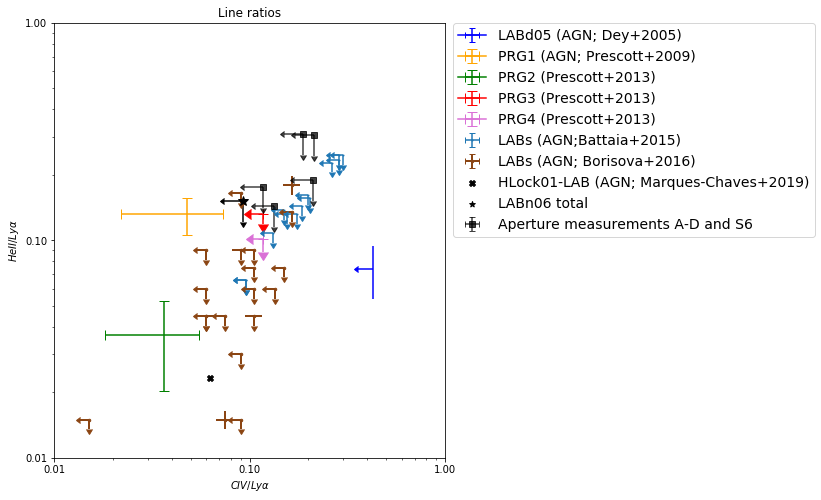}
    \caption{Line ratios for HeII, CIV, and \lya\ for the entire source (black star) along with specific regions of interest (black squares). Additionally we plot these same ratios for a sample of \lya\ nebulae in the literature \citep[][]{Dey2005, Prescott2009, Prescott2013, Battaia2015, Borisova2016, Marques-Chaves2019}, many of which are known to host an AGN (indicated in the legend). We converted all detection limits to $3\sigma$ values for this comparison. The UV emission line ratio upper limits for LABn06 are consistent with what is seen in other AGN-powered \lya\ nebulae.}
    \label{fig:diagnostic_diagram} 
\end{figure*}

In Figure~\ref{fig:diagnostic_diagram}, we show how the line ratio upper limits from the extended nebula in LABn06 compare to other \lya\ nebulae in the literature on a HeII/\lya\ vs. CIV/\lya\ diagnostic diagram. These results are consistent with the line ratio measurements and upper limits seen in other \lya\ nebulae, most of which have been identified around AGN. We note that many of the other supposedly AGN powered LAN targeted with MUSE \citep[e.g.][]{Borisova2016, Battaia2019} do not show these expected AGN emission lines yet their \lya\ halos are comparable in size and even more luminous than LABn06. While not conclusive, the UV emission line ratios found for LABn06 are consistent with the possibility of AGN-powering in this system.

\begin{figure*}
    \centering
    \includegraphics[width=\linewidth]{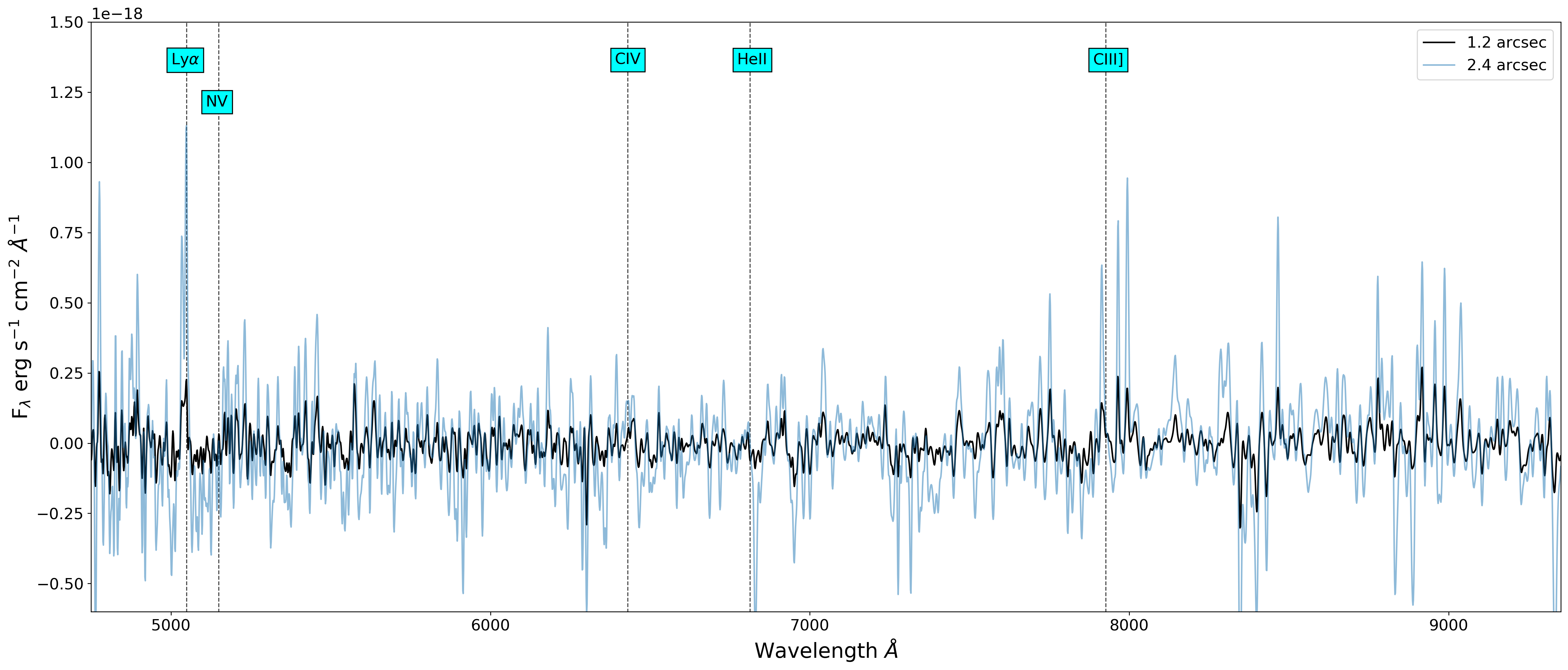}
    \caption{MUSE spectra extracted from 1.2$\arcsec$ (black) and 2.4$\arcsec$ (blue) radial apertures centered on the location of S6. Important lines typically associated with AGN are indicated by dashed vertical lines and labeled in cyan. No other emission lines besides \lya\ have a significant detection in the MUSE data.}
    \label{fig:agn_spec} 
\end{figure*}

In Figure~\ref{fig:agn_spec}, we also show the spectrum extracted from 1.2\arcsec\ and 2.4\arcsec diameter regions of the nebula centered on the location of S6, as indicated by its MIR detection. The \lya\ emission line along with other UV metal emission lines typically indicative of AGN powering are indicated by dashed black lines and cyan labels. No other emission lines besides \lya\ show a significant detection in the MUSE data.  While it would be beneficial to compare the UV metal emission upper limits obtained for S6 to those seen in typically faint or obscured AGN, the samples of AGN at z$\sim$3 for which measurements of the UV metal emission exist are 2-3 orders of magnitude brighter at 22$\micron$ than S6. Further, these samples were selected to have bright CIV$\lambda$1550 and \lya$\lambda$1216 \citep[][]{Alexandroff2013} or as some of the most optically bright AGN at z$\sim$3 \citep[][]{Borisova2016}. This suggests the need for further study into the UV emission line properties of the obscured AGN population at fainter infrared magnitudes.

\section{Discussion}
\label{sec:discussion}

Our MUSE coverage of LABn06 represents one of the deepest datacubes targeting a large \lya\ nebula. In what follows, we discuss the nature of S6, the evidence for AGN-powering and resonant scattering in this system, and 
how the morphology and kinematics of LABn06 compare to other AGN-powered nebulae. We end by revisiting the earlier arguments made for gravitational cooling in this system in light of our new data.

\subsection{The Nature of S6}
The nature of the MIR source (S6) is still not obvious. While previous investigation of the MIR SED for this source found it was best-fit by the Mrk 231 Type 2 AGN template, the FIR 70-500 $\micron$ fluxes were reminiscent of cold dust heated by star formation \citep[][]{Prescott2015b}. At the same time, \citet[][]{Prescott2015b} found that the MIR fluxes of S6 placed it solidly within  the AGN color selection boxes of \citet{Donley2012}.  However, it is true that MIR AGN color-selection can be contaminated by higher redshift (z$>$2) star-forming galaxies \citep[][]{Barmby2006, Park2010, Donley2008}. In these cases, power-law selection is an efficient way of cleanly isolating obscured AGN.  While typical star-forming galaxies tend to display a dip in emission in the MIR, AGN-heated dust is expected to produce a monotonically rising SED across the IRAC bands \citep[][]{Rees1969, Neugebauer1979, Assef2010}. For example, while a few contaminating high redshift starburst galaxies fell into the AGN selection box of \citet[][]{Donley2012} due to their measured MIR colors, the AGN nature of these sources could be ruled out due to the non-power-law shape of their IRAC SED. Taking a similar approach, inspection of the IRAC fluxes of S6 shows that its SED rises monotonically between 3.6 and 8\micron. Following the procedure in \citet[][Section 2.2]{Donley2008}, we find that its IRAC SED is well-fit by a power law, $f_{\nu}\propto\nu^{\alpha}$, with $\alpha=-1.54\pm0.17$ ($\chi^2=1.83$ $P_{\chi}=0.40$), consistent with obscured AGN power-law selection criteria.

Despite the evidence in the MIR for the AGN nature of S6, no counterpart X-ray detection was found in deep 1Ms, 2Ms, or 4Ms exposures \citep[][]{Prescott2015b} or even in more recent, 7Ms data \citep[][]{Luo2017}. Assuming S6 does contain an AGN, the lack of X-ray emission prompts us to ask whether the AGN is heavily obscured or whether it is a ramped-down source. Previous work showed that it is relatively common for a heavily obscured AGN to be undetected in X-ray observations with t$<$2Ms \citep[][]{Donley2007, LaMassa2019, Pouliasis2020}, however, it is unclear what fraction of power-law selected sources are expected to be undetected at the limit of the 7Ms data \mbox{\citep[2.7$\times$10$^{-17}$erg s$^{-1}$ cm$^{-2}$;][]{Luo2017}}. Updated studies on the X-ray properties of power-law selected AGN are needed in order to determine whether the lack of an X-ray detection in the 7Ms data can rule out the presence of an obscured AGN in S6.  A ramped-down AGN could also explain the lack of X-ray emission. In this case, the fact that we still observe AGN-powered MIR emission could indicate that the AGN ramp-down took place relatively recently. 

In deeper radio observations that now exist for this field \citep[][]{Miller2013}, we find an unresolved 5$\sigma$ 1.4GHz detection that is only 0.3\arcsec away from the location of S6. We can use this unresolved radio detection to investigate how S6 compares to the well-known FIR-radio correlation for star-forming galaxies \citep[FRC;][]{Helou1988, Yun2001, Magnelli2015}. Using a FIR SED template of the star-forming galaxy Arp 220 \citep[][]{Polletta2007}, shifted to the observed redshift of z$=$3.15 and scaled to the observed 250$\micron$ flux density of S6, we estimated the FIR (42-122 $\micron$) luminosity of S6 to be 7.69$\times$10$^{45}$erg s$^{-1}$. We K-corrected the radio luminosity assuming a power law spectral index of -0.71 for $S_{\nu}\propto\nu^{\alpha}$ \citep[][Equation 1]{Read2018} and estimated the FRC parameter to be q$_{FIR}$=2.04$\pm$0.08. This q$_{FIR}$ value places S6 within the star-forming-galaxy locus of the FRC, a region also populated by radio-quiet seyfert galaxies \citep[][]{Moric2010, Sargent2010, Padovani2011, Del-Moro2013}. This indicates that S6 does not have an excess of radio emission, making it consistent with what is expected for radio-quiet AGN and star forming galaxies.

Based on the results of the MIR AGN-selection techniques (IRAC colors and power law slope), the lack of evidence for excess radio emission, and the evidence for star formation-heated dust in the FIR, we conclude that S6 is likely a radio-quiet, obscured AGN/star-forming galaxy composite in which the AGN is either heavily obscured or recently ramped down. In what follows, we compare the LABn06/S6 system with other AGN-powered nebulae and refer to S6 as the obscured AGN.

\subsection{Morphologically \& Kinematically Centered on the AGN}

We have confirmed that this highly debated \lya\ nebulae is larger than originally reported (at least $\approx$61 pkpc in size from regions A to D, with fainter emission extending out to 173 pkpc), and is spatially offset from the brightest optically detectable galaxies in the region. Previously, \cite{Prescott2015b} provided evidence that an obscured AGN was buried $\approx$30 pkpc from the peak of the emission. In our new MUSE observations, we found that the \lya\ emission at the location of the obscured AGN is coincident in velocity with the kinematic centroid of the nebula (Figure~\ref{fig:spec_samples}, panel (b)), and that this obscured AGN lies $\approx$17 pkpc from the light-weighted centroid of the nebula, placing the AGN roughly at the center of a two-lobed \lya-emitting structure (Figures~\ref{fig:em_significance} and ~\ref{fig:NBSB}). The spatial coincidence may be even closer if the gas in the western lobe is distinct from LABn06 as suggested by the steep velocity gradient in this region. To explore this possibility, we perform the following exercise: we remove the gas in the western lobe from our calculation of the light-weighted centroid of the \lya\ emission by masking out all values West of the surface brightness dip (vertical dashed line) in Figure~\ref{fig:spec_samples} panel (a). In this case, we find that the light-weighted centroid lies even closer, within 0.3\arcsec\ (2.3 pkpc) of the AGN. Taken together, the MUSE data show that the AGN is likely located at both the spatial and kinematic center of the \lya\ nebula.

While the spatial proximity of the obscured AGN to the light-weighted centroid of the \lya\ emission could be a coincidence, we think this scenario is much less likely. The obscured AGN was first identified using standard Spitzer/IRAC MIR color selection \citep[e.g.,][]{Lacy2004, Lacy2007,Donley2012}. Using the compilation of \citet[][their Figure 2]{Mendez2013} and extrapolating the \citet[][]{Donley2012} curve to the flux density of the obscured AGN in LABn06 ($\approx$1.57$\times10^{-29}$ erg s$^{-1}$ cm$^{-2}$ Hz$^{-1}$), we estimate that the expected surface density of such obscured AGN is $\approx$10$^3$-10$^4$ degree$^{-2}$ dex$^{-1}$. In LABn06, the obscured AGN is located within a radius of 2.2\arcsec\ (0.3\arcsec) of the light-weighted centroid of the nebula, corresponding to a circular area of $\approx$1$\times10^{-6}$ degrees$^2$ ($\approx$2$\times10^{-8}$ degrees$^2$). Therefore, by multiplying this area by the expected surface density, we estimate that there is a 0.1-1\% (0.002-0.02\%) chance of an obscured AGN (as selected by \citet[][]{Donley2012} color-cuts) showing up by chance within 2.2\arcsec\ (0.3\arcsec) of the light-weighted centroid of LABn06.

Our deep MUSE data provide indirect evidence for AGN-powering in LABn06, with the nebula being morphologically and kinematically centered on an obscured AGN, and with derived upper limits on HeII and CIV that are consistent with AGN-selected systems. In what follows, we explore how LABn06 compares to other AGN-powered nebulae, both in terms of morphology and kinematics.

\subsection{Morphology compared to other AGN-powered Nebulae}

In \citet[][]{Borisova2016}, the authors targeted bright quasars using MUSE and found extended \lya\ emission around every single source. These presumably AGN-powered \lya\ nebulae span a range of sizes (100-300 pkpc) and luminosities (10$^{43.3}$-10$^{44.6}$ erg s$^{-1}$). Compared to this sample, LABn06, with a full extent of $\approx$173 pkpc and a \lya\ luminosity of $\approx$10$^{43}$ erg s$^{-1}$, seems to be comparable in size, although only half as luminous. Although rigorous comparisons should use size measurements down to the same surface brightness threshold \citep{Wisotzki2018, Leclercq2020}, this qualitative comparison suggests that LABn06 is quite extended for its low surface brightness and likely falls below the correlation implied for other AGN-powered \lya\ nebulae at this redshift \citep[][]{Christensen2006}.

To make a more robust comparison of the nebula size, we use the best-fit scale radius of the circularly averaged surface brightness profile. LABn06 is best described by an exponential with a scale radius of 24.2 pkpc. In Figure~\ref{fig:SBprofiles} panel (b), we plot the profiles for LABn06 and eight other sources best-fit by an exponential profile, all normalized to 25 pkpc. LABn06 appears most similar to the Lyman Break Galaxy stack of \citet{Steidel2011}, both in terms of scale radius (25.2 pkpc) and in terms of profile shape at distances between 25-60 pkpc. In the outer regions, LABn06 also resembles other AGN-powered nebulae observed with MUSE in terms of profile shape, although it is a bit smaller in terms of scale radius. LABn06's scale radius is at the lower end of the range spanned by the 5/19 sources from \citet[][r$_{scale}$ = 20.69, 29.71, 38.88, 40.12, 55.66 pkpc]{Borisova2016} that were also best-fit by an exponential profile. 

LABn06 shows a similar level of asymmetry compared to other AGN-powered nebulae, particularly those around Type II AGN. The dimensionless asymmetry parameter ($\alpha$) calculated for LABn06 is roughly at the median $\alpha$ value of the population of Type II AGN investigated by \citet[][]{denBrok2020}. 
The asymmetry profiles derived using Fourier decomposition in Figure~\ref{fig:FDecomp} revealed that while LABn06 appears relatively symmetric at intermediate radii, it exceeds the typical asymmetry of Type I AGN at both small ($<$10 pkpc) and large ($>$50 pkpc) distances from the AGN. This fact combined with the bipolar morphology, and the relatively mild velocities are all consistent with LABn06 being only moderately inclined along our line-of-sight, potentially giving us an ``out-of-the-cone'' view of the system. 

Despite the similarities to other AGN-powered systems, at small radii LABn06 is unique in showing a pronounced hole around the position of the obscured AGN. This could be due to high central HI column density and/or dust obscuration preventing \lya\ photons from escaping the nebula in the vicinity of the obscured AGN. Alternatively, it could be due to a lack of HI gas in the central region. Finally, it is possible that this dip reflects a recent ramp down in AGN power roughly 10$^5$ years ago, in which case, we are seeing an ionization echo in the outskirts of LABn06.  

More speculatively, if the AGN in LABn06 has recently ramped down, the fact that one lobe of the nebula is brighter than the other might suggest that we could be viewing a late-day version of the \lya\ nebula in \citet[][]{Weidinger2005}, which showed a biconical structure inclined at an angle around a z$\sim$3 quasar. If the central engine shuts off, the light traveling to the observer from the backside cone would be delayed relative to that from the front side cone. This might create enough of a time window that we would see a bicone that is not symmetrically illuminated. If this interpretation is correct for LABn06, it would imply that the northern lobe in this system is further away from us along the line-of-sight than the southern lobe.

While to our knowledge this is the first case of a \lya\ nebula with a central hole at the location of the AGN, we note that having an AGN offset from the brightest regions of \lya\ emission is not unprecedented. Other examples of offset AGN powering extended \lya\ nebulae are known to exist \citep[][]{Kurk2002, Prescott2009, Prescott2012b}. In addition, hydrodynamical simulations of \lya\ nebulae reveal that the brightest regions can be offset substantially from the location of star forming galaxies if AGN are included as a powering mechanism \citep[][]{Kimock2020}.

\subsection{Kinematics compared to other AGN-powered Nebulae}
AGN-powered nebulae in the literature show a diversity of kinematics. For example, \cite{Borisova2016} found that most of the nebulae showed relatively chaotic velocity fields, without any coherent gradient in velocity. However, the two largest nebulae presented clear velocity shear, in one case along the major axis of the nebula but in the other, surprisingly, along the minor axis. Similarly, \citet[][]{Prescott2015a} found signs of large scale rotation as indicated by \lya, CIV, HeII, and CIII] in the \lya\ nebula PRG1 at z$\sim$1.7. \citet[][]{Zafar2011} identified evidence for ordered motion in the \lya\ nebula around the binary quasar Q0151+048 at z$\sim$1.9, and \citet[][]{Battaia2019} identified an obvious velocity gradient centered on two LAE and two AGN in an extended \lya\ nebula at z$\sim$3.2. Finally, \citet[][]{Herenz2020} found clear signs of a velocity gradient perpendicular to the principal axis of LAB1 at z$\sim$3.
In this context, LABn06 is well within the range of other AGN-powered nebulae, with line kinematics that are somewhat chaotic, but with some signs of coherent velocity shear along the major axis, as seen in Figure \ref{fig:kinematics}.

In terms of velocity dispersion, other \lya\ nebulae in the literature show a wide range of values. For their sample of z$\sim$3 \lya-emitting nebulae, \citet[][]{Borisova2016} found velocity dispersions of $\sigma_v$=150-450 km s$^{-1}$, although the three largest sources and the two radio-loud sources displayed a peak in velocity dispersion near the location of their AGN. Spectroscopically derived line profiles for these nebulae show a range of both narrow and broad lines with single, asymmetric shapes or, in a few sources, hints of double peaks. Similarly in LAB1, \citet[][]{Herenz2020} found a range of velocity dispersions ($\sigma_v$=200-550 km s$^{-1}$), although the highest dispersion was found in the region occupied by identified sources. The line profile derived from this region of highest dispersion appeared to be double peaked, although another double peaked profile was found in a region devoid of sources.

In LABn06, we see varying \lya\ line profiles across the nebula (Figure~\ref{fig:spec_samples}) and apparent velocity dispersions of 
$\sigma_v\approx$350-450 km s$^{-1}$. The apparent velocity dispersions do reach $\sigma_v\approx$500 km s$^{-1}$ to the North of the obscured AGN's MIR position. However, from Figure \ref{fig:SBprofiles} panel (b) we can see that regions of highest apparent velocity dispersion tend to show double peaked profiles. Thus, these high $\sigma_v$ values likely reflect more complicated line profile shapes rather than more chaotic kinematics. Complicated, double-peaked line profiles are expected from \lya\ radiative transfer calculations,
as \lya\ photons produced at line center need to be either red- or blue-shifted in order to escape the core of the profile \citep{Verhamme2006}. While it is difficult to draw strong conclusions in the case of LABn06, the double peaked profiles suggest that resonant scattering of \lya\ photons is important in this system, and that in some regions additional kinematics may be responsible for producing asymmetric line profiles. The conclusion that resonant scattering appears to be important and that the obscured AGN is the dominant power source together make a clear prediction that this nebula should show rising \lya\ polarization as a function of radius from the AGN location \citep[e.g.,][]{Rybicki1999, Hayes2011}.
It would therefore be interesting to obtain imaging polarimetry  or spectro-polarimetric data to confirm both of these claims.

It is possible that the gas emitting from the biconical structure might have been part of, or impinged upon by, a galactic scale wind driven by massive stars \citep[e.g.,][]{Wilman2005} or an AGN-powered radio jet \citep[e.g., ][]{Swinbank2015}. In modeling the evolution of \lya\ nebulae in galaxy formation simulations, \cite{Kimock2020} found that $\sim50\%$ of the gas in the halo of a simulated nebula has crossed the virial radius of the central galaxy one time by $z=2$, meaning it was likely part of an outflow (their Figure 17). While the stellar wind scenario will probably be less likely on such large scales, deeper low-frequency radio imaging of this region would be needed to rule out the possibility of a past interaction between a now fading radio jet and the emitting nebula. 

\subsection{Gravitational Cooling, Revisited}
LABn06 has often been cited as an example of gravitational cooling in \lya-emitting nebulae, but additional data continues to call this into question. Most of the \lya\ emission seen in LABn06 is offset substantially from star-forming galaxies in the region and is relatively clumpy within a two-lobed structure. By contrast, \lya\ nebulae that are lit up via cooling radiation in hydrodynamical simulations are centered on the region of densest star formation \citep{Rosdahl2012}, and appear smoother and less concentrated than \lya\ nebulae that host AGN \citep{Kimock2020}. The presence of a ``hole'' in the center of the \lya\ nebula is particularly at odds with theoretical expectations for gravitational cooling. Furthermore, gravitational cooling nebulae should be associated with the most massive halos \citep{Ao2020}, but even in cases massive enough to produce gravitational cooling radiation, \lya\ emission driven by AGN fluorescence is still expected to dominate over the lower surface brightness cooling emission \citep{Kimock2020}.

\section{Conclusion}
\label{sec:conclusion}
The vast majority of \lya\ nebulae targeted with wide-field IFS instruments to date have been selected due to the presence of a known AGN. Here, we used the MUSE instrument aboard the VLT's 8.2m telescope to simultaneously image and obtain spectroscopic coverage of a \lya-selected nebula at $z\sim3.2$, finding that:

\begin{itemize}
    \item The source originally identified as Source 6 is likely a radio-quiet, obscured AGN/star-forming composite galaxy.
    \item The location of the obscured AGN lies within 17 pkpc of the light-weighted centroid of the \lya\ emission, and the gas surrounding the AGN is at the kinematic center of the nebula. The nebula therefore appears to be morphologically and kinematically centered on the obscured AGN.
    \item The full extent of LABn06 is $\approx$173 pkpc, measured down to a surface brightness of 4.0$\times10^{-19}$ erg s$^{-1}$ cm$^{-2}$ arcsec$^{-2}$. At large radii ($\gtrsim$25 kpc), the \lya\ nebula surface brightness profile is well-fit with an exponential, resembling the profiles of other AGN-powered nebulae and stacked Lyman-break galaxies. At small radii, however, LABn06 shows an unusual dip in its surface brightness profile, with the brightest peak offset spatially from the location of the AGN.
    \item LABn06 shows a level of asymmetry in between the nebulae seen around Type I and Type II AGN. In terms of second-order moments, LABn06 is quantified as being circularly asymmetric, similar to Type II AGN, and in terms of a Fourier decomposition approach, it shows circular symmetry at moderate radial distances but a relatively asymmetric profile compared to the majority of Type I AGN at larger radial distances. 
    \item The kinematics of the nebula show signs of coherent motion and a range of apparent line widths up to $\sigma_v\approx$500 km s$^{-1}$, similar to what has been observed around other AGN-powered nebulae observed with MUSE.
    \item The \lya\ emission lines at different locations across the nebula show double peaks and asymmetric profiles, suggesting that resonant scattering is playing a role in the system.
    \item AGN-powering cannot be conclusively demonstrated based on high ionization lines, but the constraints on the CIV/\lya\ and HeII/\lya\ ratios measured within the nebula are consistent with measurements from other AGN-powered \lya\ nebulae.  
\end{itemize}

These results are consistent with the obscured AGN being largely responsible for powering the extended \lya\ emission in LABn06. Further observations will be needed to understand the central dip in the surface brightness profile, and whether this indicates an unusual distribution of gas or dust in this system or time-variability in the ionizing output from the AGN. Additionally, obtaining \lya\ polarimetry to constrain the effects of \lya\ photon scattering and low-frequency radio observations to rule out the presence of past radio-jet interactions with the gas will greatly benefit our understanding of this source. 

\newpage

\section{Acknowledgements}
The authors would like to thank Sandra Raimundo for assistance with the MUSE DRS pipeline, Jakob den Brok for providing comparison data for our analysis of the asymmetry in LABn06, and Christian Herenz for introducing the use of LSDcat. We would also like to thank Kristian Finlator, Audrey Dijeau, and Natalie Wells for helpful discussions and the anonymous referee for suggestions that improved the quality of this work. 

This material is based upon work supported by the National Science Foundation under Grant No. AST-1813016.

This research has made use of the services of the ESO Science Archive Facility. Based on observations made with ESO Telescopes at the La Silla Paranal Observatory under programme ID 0102.A-0484(A) and ID 106.21GX.001.

This work is based on observations taken by the 3D-HST Treasury Program (GO 12177 and 12328) and on observations taken by the CANDELS Multi-Cycle Treasury Program with the NASA/ESA HST, which is operated by the Association of Universities for Research in Astronomy, Inc., under NASA contract NAS5-26555. The Cosmic Dawn Center of Excellence is funded by the Danish National Research Foundation under the grant No. 140.

\bibliography{sample63}{}
\bibliographystyle{aasjournal}

\end{document}